# Tuning the mechanical properties of organophilic clay dispersions: particle composition and preshear history effects

*Nikolaos A. Burger[1,2], Benoit Loppinet[1], Andrew Clarke[3] and George Petekidis[1,2]*

[1] *IESL-FORTH, P.O. Box 1527, GR-711 10 Heraklion, Greece*
[2] *Department of Materials Science & Technology, University of Crete, Heraklion 70013, Greece*
[3] *SLB, Schlumberger Cambridge Research Ltd., High Cross, Madingley Road, Cambridge CB3 0EL, UK*

**Abstract**

Clay minerals are abundant natural materials used widely in coatings, construction materials, ceramics, as well as being a component of drilling fluids. Here, we present the effect of steady and oscillatory preshear on organophilic modified clay gels in synthetic oil. Both platelet and needle-like particles are used as viscosifiers in drilling fluid formulations. For both particles the plateau modulus exhibits a similar concentration dependence, $G_P \sim c^{3.9}$, whereas the yield strain is $\gamma_y \sim c^{-1}$ for the platelets and $\gamma_y \sim c^{-1.7}$ for the needles. Mixtures of the two follow an intermediate behavior: at low concentrations their elasticity and yield strain follows that of needle particles while at higher concentrations it exhibits a weaker power law dependence. Furthermore, upon varying the preshear history, the gel viscoelastic properties can be significantly tuned. At lower (higher) clay concentrations, preshear at specific oscillatory strain amplitudes or steady shear rates, may induce a hardening (softening) of the dispersions and, at all concentrations, a lowering of the shear strain. Hence, in needle dispersions preshear resulted in changes in the volume fraction dependence of the elastic modulus from $G_P \sim c^{3.9}$ to $G_P \sim c^{2.5}$ and of the yield strain from $\gamma_y \sim c^{-1.7}$ to $\gamma_y \sim c^{-1}$. However, small angle X-ray scattering showed not much structural changes, within the q-range covered. Our findings indicate ways to design colloidal organoclay dispersions with a mechanical response that can be tuned at will.

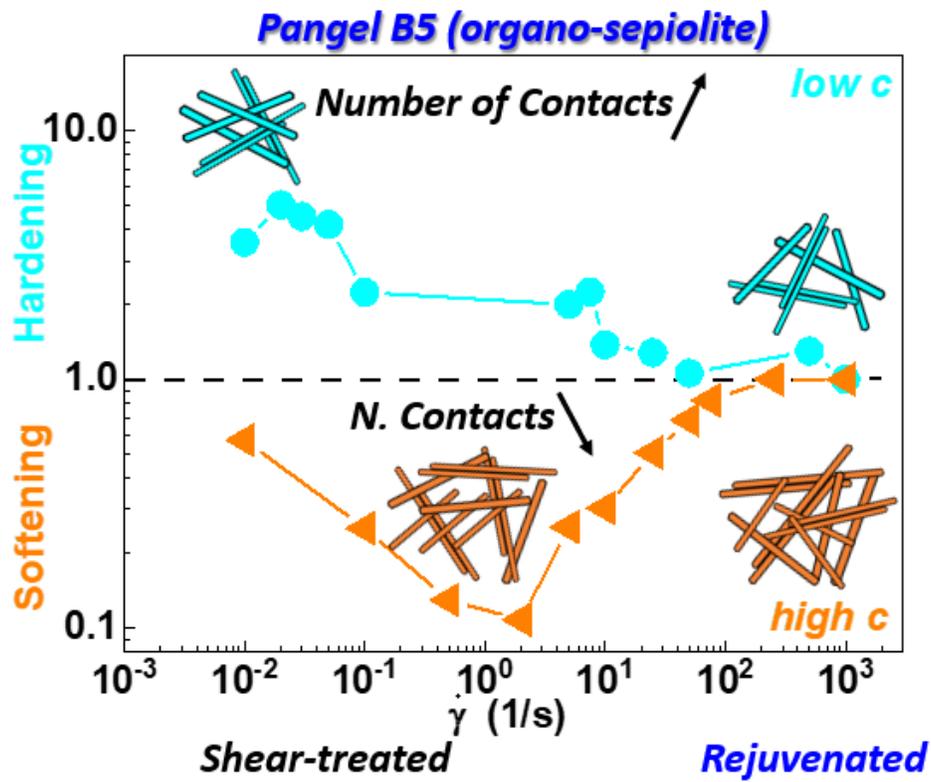

Graphical Abstract: Variation of plateau modulus of Pangel B5 dispersions through shear treatment. Tentative cartoon of shear effect on needle dispersion: depending on clay concentration *(lo*w -cyan, high- orange*)* shear treatment may increase or decrease the number of contacts between particles.

# I. INTRODUCTION

Colloidal gels are out-of-equilibrium disordered phases of colloidal dispersions that are ubiquitous both in nature and in industrial and consumer products [1–4]. Gelation evolves through two distinct pathways. At intermediate volume fractions colloidal gels can be formed by arrested phase separation. In that case, the non-linear response is described by a double yielding behavior[5–7]. On the other hand, at low volume fractions fractal gels, are formed by an irreversible aggregation at significant levels of inter-particle attraction strength [8]. Homogenous aggregation of attractive sticky particles evolves with time until a space-spanning (macroscopic) network is formed [9,10]. Such mesoscale structures are characterized by (at least) two distinct length scales; small length-scales relevant to the radius of individual particles and the range of attractions, and longer length-scales relating to the size of aggregates (clusters). These aggregates are connected to each other by intermolecular forces$(> k_BT)$ and form with a floc of size, $(\xi)$, with the respective volume fraction, $\varphi(\xi)$, which scales with the fractal dimension, $(d_f)$ of the respective flocs. Diffusion limited cluster aggregation (DLCA) and the reaction limited cluster aggregation (RLCA) at low volume fractions lead to different internal aggregate structures represented by different fractal dimension $d_f$ (ranging from 1.7 to 2.3). Both mechanisms have been explored experimentally and via computer simulation [11–15]. The nonlinear mechanical response of such mesoscale structures is characterized by a single yield process but, in addition, exhibit a rich phase behavior with an intriguing rheological behavior with complex interrelated and antagonistic phenomena such as shear thinning, shear induced yielding and structural break up, aging and thixotropy with accompanied by a broad level of timescales [4,16–21]. The concentration dependence of elastic modulus and yielding is an important rheological characterization which we expect should reflect the type of aggregation. The structure-elasticity relationship is not straightforward and depends by many parameters (attraction strength, volume fraction, surface etc.) [22,23]. Early studies have indicated a strong link between rheology and the internal structure and microscopic dynamics in such heterogeneous systems [24–26]. This connection is more challenging when we try to characterize poly-disperse dispersions of Brownian and non-Brownian particles and clusters. The organoclays used here are such cases of mixtures that also find tremendous use in the oil & gas industry and are components of drilling fluid formulations [27–30]. By design, the formulations should possess a pseudo-plastic response with a yield stress of the order of a few Pascals [31–34]. Whereas the formulation is complex, providing a yield-stress is the main task of clay particles. In other applications, organoclay dispersions play a vital role as adsorbents of organic pollutants and as colloidal particles in polymer-colloid nanocomposites [35–38]. Because of their out-of-equilibrium nature; the mechanical properties are expected to be process dependent. Shear is not only a relevant external parameter for industrial applications, but is also a very efficient way to re-disperse and 'rejuvenate' samples through breaking of existing clusters (aggregates) [39–43]. When a metastable system (such as colloidal gel) is subjected to shear forces significantly higher than the existing particle level thermal or interparticle forces, the system can be driven to metastable states. Those states are otherwise impossible to reach due to the high energetic barriers that are insurmountable at rest. Therefore, shear history is an external parameter that directly impacts the structure and mechanical properties of colloidal dispersions which is both interesting from a fundamental point of view and important for industrial purposes [44–52].

Here, we focus on organo-clay particles of varying shape (platelet morphology VG-69™ organophilic clay (mark of SLB) and needle morphology Pangel B5 ® (registered trademark of Tonga S.A.)), their interactions and their respective gels. We study the effect of shear history on the rheological properties and we identify reliable ways to tune their elastic modulus. Interestingly, we observed that shear treatment also affected yield stress and yield strain. We first provide structural information of the chemically modified clay particles used here and then discuss our rheological findings. For all the dispersions, preshear induced hardening is observed at lower concentrations whereas hysteresis and softening is detected at intermediate to higher concentrations. To probe these observations, we designed a specific shear protocol to follow the gel evolution upon cessation of shear flow. For the individual clays, the plateau modulus and yield stress exhibit a distinct power-law increase with concentration ($G_P \sim c^{3.9}$ and $\sigma_y \sim c^{2.9}$) over the whole concentration regime studied following model predictions for fractal colloidal gels. In contrast, the yield strain appeared to have a much weaker dependence on concentration $\gamma_y \sim c^{-1}$ for VG-69 platelets and $\gamma_y \sim c^{-1.7}$ for Pangel B5 needles. When we mix the two components by 50-50 wt. %, we find an interesting emergent behavior: At low total clay concentrations the dispersion has linear and nonlinear properties (elasticity, yield strain) similar to the Pangel B5 needles alone, whereas at higher concentrations the presence of VG-69 platelets significantly affects the behavior. Therefore, in contrast to individual clays, mixtures have a weaker power–law behavior of elasticity and yielding ($G_P \sim c^{2.8}$, $\gamma_y \sim c^{-0.4}$) which suggest significant changes in the connectivity and weakly connected flocs. To elaborate on our findings regarding shear induced tuning, we combine shear rheology with SAXS to reveal a weak shear-induced anisotropy upon steady shear flow which relaxes immediately upon cessation of flow. Moreover, from the azimuthal average (for the isotropic patterns) only minor changes were observed during oscillatory shear.

## II. MATERIALS AND METHODS

### A. Materials

In this study we use two organophilic clay gels, where the primary particles differ in shape. Such clay particles are the viscosifying components of full drilling fluid formulations used frequently in the oil and gas industry. The hydrophobically modified clays comprise VG-69 platelets, an amine-treated bentonite, disk-like particle, and Pangel B5 needles, an organophilic sepiolite with needle-like primary particles [53]. Clay particles have a refractive index (RI) of approximately $RI_{clay} \approx 1.51$, with the oil solvent having an $RI_{oil} \approx 1.447$, at 25 °C [34].

### B. Materials preparation and characterization

The dispersions were prepared using the hydrophobically modified clays VG-69 platelets and Pangel B5 needles as received. In the case of mixtures, the individual dispersions are prepared and subsequently mixed. Since different sample preparation protocols (temperature, shear treatment, preparation time) can induce significant changes in material properties, the dispersions were prepared using the following specific protocol: we first add the mineral oil, Clairsol 370, a mixture of organic apolar solvents together with Milli-Q water (5 $wt\%$ of the dispersion clay content, which at our maximal clay composition is approximately the water solubility limit) that will activate the clay particles during mixing and shear to fully disperse the water at a molecular level, then the clay is gradually added and is mixed with a Dremel (equivalent with a high-shear mixer, say a Silverson L4RT) at 6000 rpm for at least 25 minutes. The samples were prepared at room temperature (~20ºC), whereas during the high-shear procedure the temperature was allowed to rise to about 45ºC. To ensure homogenous dispersion without air bubbles prior to measurement, samples were placed overnight on a bottle roller.

Scanning electron microscopy (SEM) imaging was performed with a JEOL JSM 6390LV scanning electron microscope operated at 15 $to$ 20 $kV$. Before sample imaging, all samples were dried and sputter-coated with ca. 10 $nm$ of gold (Au).

***Rheometry:***

We used two stress-controlled rheometers (MCR-501 and MCR-302, Anton-Paar, Austria), operating both in stress and strain-controlled mode. We utilized a home-made plastic serrated cone - plate geometry ($d = 40\ mm$, cone angle $= 2.54°$ and truncation $= 0.125\ mm$), and a home-made cover to eliminate solvent evaporation, even though Clairsol 370 has a high boiling point ($T > 280\ °C$). To verify our findings we also performed additional measurements with other geometries, i.e. with a smooth parallel plate ($d = 50\ mm$) and a sandblasted cone plate ($d = 40\ mm$, $cone\ angle = 1.02°$ and $truncation = 0.072\ mm$), to quantify possible slip effects. The samples were always loaded at 25 °C. To avoid slip effects we also used a rough bottom plastic plate. The rheometers are equipped with a Peltier unit for temperature control ($\pm 0.1\ °C$). Samples were always shear rejuvenated at $1000\ s^{-1}$ for 1 minute prior to any rheometric measurement. Such preshear is sufficient to reset any structural evolution and results in a reproducible initial state (for more details see below and in the S.I.). Rheo – SAXS measurements were performed in a Taylor – Couette cell composed of a bob of radius $10\ mm$, a cup of radius $11\ mm$ and a height of $40\ mm$) connected to a stress-controlled rheometer (Haake RS6000, Thermo Scientific) [54].

**III. EXPERIMENTAL RESULTS**

**A. Structural characterization of clay particles**

In Fig. 1(A), SEM images of the raw powders show the individual shape and the high polydispersity of the clay particles. We point out that the clusters visible in the images are expected to be both dispersed and exfoliated to some extent during the high shear preparation of the suspension. The needles comprising Pangel B5 have an average radius of about $50\ nm$ and length of about $1\ \mu m$, i.e. exhibit a high aspect ratio. The platelets comprising VG-69 organophilic clay appear to consist of plates with thicknesses of the order of $1\ nm$. Further detail of chemical composition of these clays can be found in [55,56]. The different magnifications allow us to visualize the different length scales present, i.e. individual needles or platelets at high magnification and larger aggregates and clusters at lower magnification. Since we expect the preparation to breakdown these clusters, we also expect most of the mass to be in the form of small platelets or needles. In general, organoclays are considered to be well exfoliated under such preparation conditions. We expect both dispersions to consist of well exfoliated stacks with a small average number of sheets per stack, $< N \leq 2 >$ [57,58].

Our understanding of disorder gel structure (with no specific particles orientation) is that dispersed stacks form flocs of particles connected through weak attraction (van der Waals and Hydrogen bonding) eventually percolating and forming an elastic network with isotropic distribution of orientation. This is in agreement with other studies in particular cryo-SEM pictures from similar samples showing very open network of polydisperse needles or plates packed together [59].

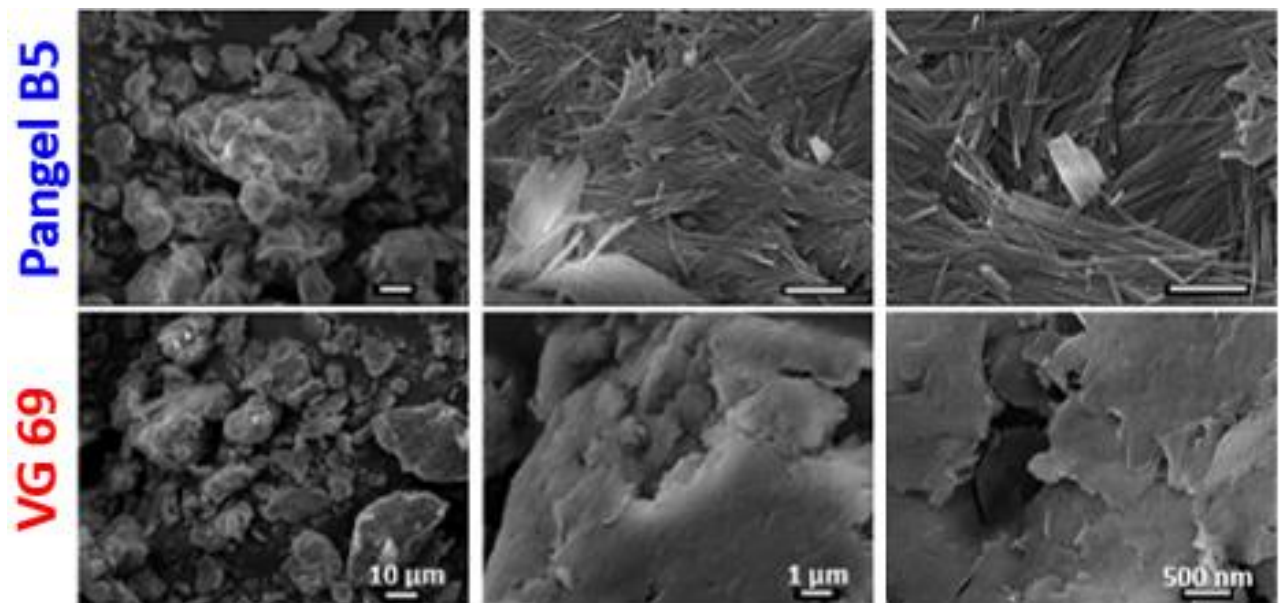

**Fig. 1:** SEM images of Pangel B5 needles (above) and VG-69 platelets (below) dry raw powders at different magnifications as indicated by scale bars at $10\ \mu m, 1\ \mu m$ and $500\ nm$.

**B. Linear rheology at different concentrations of organo-clay particles**

As previously discussed, measurement of rheological properties of such complex materials can be challenging. To produce consistent reproducible data, rheological measurements need to be performed following strict protocols. Thus, we adopted the following protocol: Prior to a dynamic frequency sweep (DFS) we perform: i) preshear at $1000\ s^{-1}$, for $60\ s$ which is followed by a resting period of $t_w = 1200\ s$. During this waiting period the evolution of $G'$, $G''$ is monitored by small amplitude oscillatory shear (SAOS) measurements at $\omega = 1\ rad/s$ and $\gamma = 0.1\ \%$. Note, the effect of preshear amplitude and waiting time ($t_w$) duration was explored and the results can be seen in Fig. S1. We observed that all dispersions studied (independent of the type or concentration of the clay particles) undergo a fast (~1 s) gel reformation (liquid to gel transition) after flow cessation. This is an important observation and should be related to the tuning properties of such dispersions. The plateau modulus of the dispersions evolves slowly with time for the first two hours following a sublinear increase as $G'\sim t_w^{0.3}$ and then approaches a steady state with a very weak (slow) time evolution at longer times (as $G'\sim t_w^{0.15}$) as can be seen in Fig. S2. We refer to the latter process as slow aging. For concentrations higher than $0.5\ wt\%$, all samples exhibit a solid-like response with $G' > G''$ with $G'$ almost frequency independent and $tan\delta \sim 0.1$ for the whole frequency range.

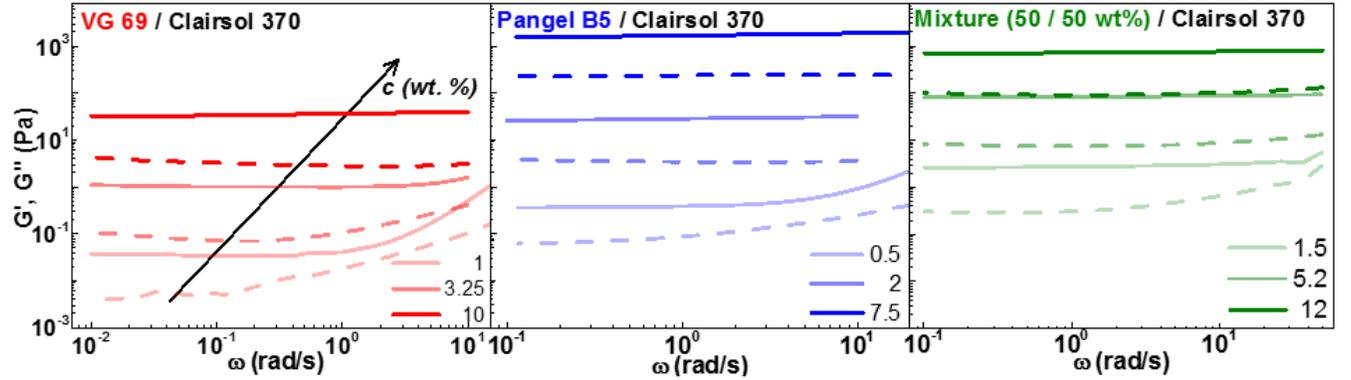

**Fig. 2:** Storage ($G'$, lines) and Loss ($G''$, dashed lines) modulus (Pa) as a function of frequency at different clay concentration at $\gamma = 0.1\%$ for VG-69 platelets (left), Pangel B5 needles (middle) and their mixtures (right plot) dispersed in Clairsol 370. Measurements are performed from high to low frequencies. The black arrow indicates increasing clay concentration.

The frequency dependent storage ($G'$) and loss ($G''$) modulus at different concentrations of the individual clays, and their mixtures (50 /50 $wt\%$) are shown in Fig. 2. At the same clay concentration, Pangel B5 dispersions form significantly stronger gels compared to VG-69 dispersions. In addition, Pangel B5 dispersions can form a gel at lower concentrations. In Fig. 3 we show the dependence of the gel storage modulus with clay concentration (mass fraction) for each of the three systems. The gel points (onset of gel formation) are indicated with the blue, green and red lines for each of the three systems respectively. Although the VG-69 dispersions are significantly weaker (red stars), they appear to exhibit the same power law behavior as for the Pangel B5 dispersions (blue squares) with both following an increase of the elastic modulus as $G_P \sim c^{3.9}$. This scaling exponent is similar to those reported for many other colloidal gels at low volume fractions stemming from the fractal character of the network and its corresponding fractal dimension [57,60,61]. This will be analyzed and discussed in detail below. In contrast, mixtures (50/50 wt. %) (Green circles) appear to have a very interesting behavior: At low concentrations the elasticity of the mixture is dominated by the network of the needles whereas at higher concentrations ($\sim 2\ wt.\%$) there is a significantly weaker plateau modulus, likely caused by a disturbance of the Pangel B5 needle network by the VG-69 platelets. Furthermore, we observe a strong discrepancy in the scaling law of the elastic modulus and yield strain which exhibit a weaker concentration dependence, namely $G_P \sim c^{2.8}$ and $\gamma_y \sim c^{-0.4}$ (see Fig. S6). Most importantly, discussed further below, mixtures have a surprisingly different yielding behavior with increasing concentration compared to the pure clay systems which renders the origin of elasticity in the pure clay gels questionable. This rheological behavior is however, in-line with previous studies demonstrating that platelet particles may disrupt fiber-based networks with an impact on their rheology (yield stress, linear elasticity, yield strain) and structure [62–67].

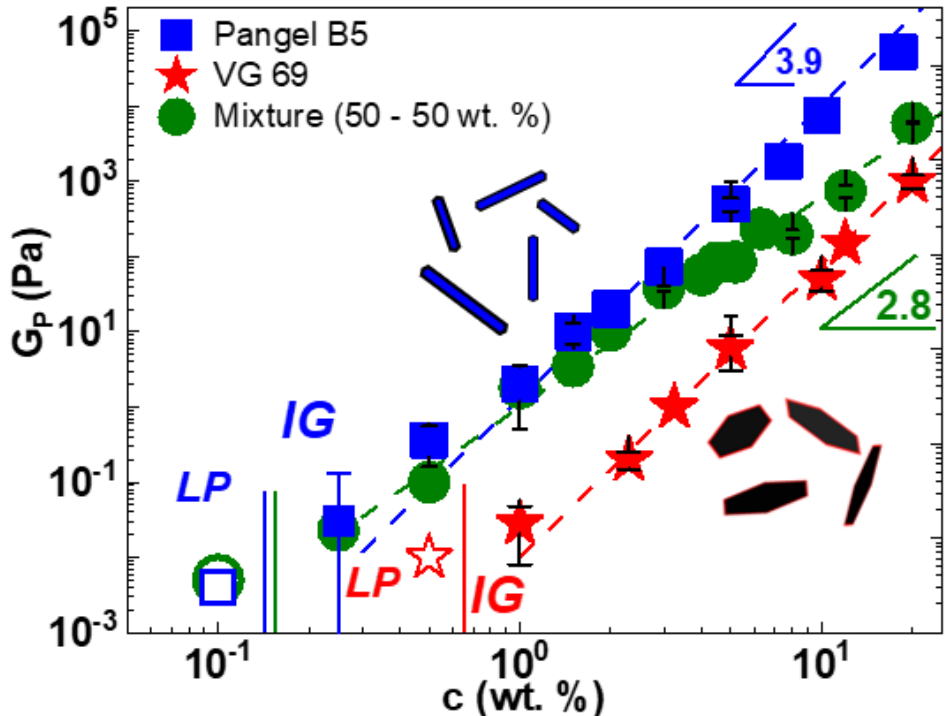

**Fig. 3:** Evolution of plateau modulus ($G_P$) as a function of clay concentration (measured at $\omega = 1\ rad/s$, and $\gamma = 1\%$) for Pangel B5 dispersions (blue squares – blue rods), VG-69 dispersions (red stars – platelets) and mixtures (green circles). Data derived from DFS measurements and the corresponding curves depicted in Fig. 2. Open symbols indicate liquid phase (LP) and filled symbols, isotropic gel (IG).

**C. Hysteresis and elasticity map of organoclays**

In this section we report the dynamic yielding behavior of the clay dispersions as measured through oscillatory rheometry. A fixed frequency is chosen ($\omega = 1\ rad\ s^{-1}$) with strain amplitudes between $\gamma = 0.1 – 100\%$ (increasing and decreasing). The measurements are performed after the rejuvenation protocol described previously. In Fig. 4 we show as an example results for a low concentration of Pangel B5 needles. Blue circles correspond to increasing and light blue stars to decreasing strain amplitudes. Vertical solid arrows indicate the cross-over of $G'$ and $G''$ whereas horizontal dashed arrows indicate the value of $G'$ at $\gamma = 0.1\%$ (i.e. the linear response value) of rejuvenated (i.e. presheared and rested and before strain amplitude increase), $G'_{Rej}$ and dynamic strain sweep (DSS) treated (after the strain amplitude increase and subsequent decrease), $G'_{DSS}$. Upon strain amplitude increase we observe a strain hardening behavior for the Pangel B5 sample. The elastic modulus was found to grow from its initial, linear regime value (see Fig. 4) for strain amplitudes increasing from $\gamma \sim 3\ to\ 30\%$. This increasing elasticity is found to be frequency independent, at least for the frequency regime probed here (i.e. $0.1 \leq \omega \leq 10\ rad/s$, see Fig. S4), and leads to a $G'$ peak (in addition to a strong $G''$ peak) which is subsequently followed by the usual $G', G''$ decrease as observed in wide range of yield stress fluids under application of constant with time oscillatory strain or stress in the non-linear regime, known as large amplitude oscillatory shear (LAOS). We should note that SAXS experiments under DSS have not revealed any major structural changes in the system in the probed $q$ range, and only a tiny change of the scattering intensity $I$ in the low $q$ regime is apparent (see Fig. S5). So, whatever structural change has occurred there is a permanent and significant increase in $G', G''$. Hence the observed small structural changes induce significant changes in the elasticity.

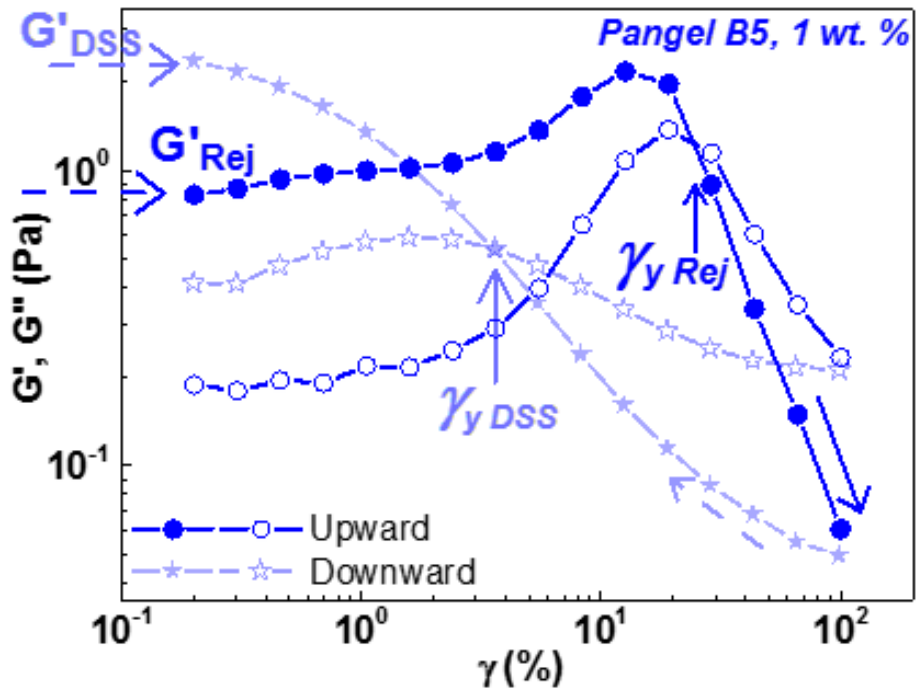

**Fig. 4:** Dynamic strain sweep measurements performed at $\omega = 1\ rad/s$ from: low to high (blue circles) and high to low (light blue stars), as also indicated by the black arrows. Filled symbols $G'$, open symbols $G''$. Solid arrows indicate the strain at crossover of $G'$ and $G''$ of the rejuvenated, $\gamma_{y\ Rej}$, and the DSS treated, $\gamma_{y\ DSS}$, dispersion. Dashed arrows indicate $G'$ at $\gamma = 0.1\%$ for the rejuvenated, $G'_{Rej}$, and DSS treated, $G'_{DSS}$, dispersion.

Colloidal gels are well-known systems that undergo significant hysteresis in their viscoelasticity upon shear rate changes (step up or step down) as well as more profound liquid to solid (gel) transition upon flow cessation. The former is related to multiscale changes in the microstructure and the bond connectivity under shear and the subsequent relaxation upon flow cessation [68,69]. Similar hysteresis is detected in the $G'$ and $G''$ curves upon strain amplitude increase and subsequent decrease as shown in Fig. 4. There we see that upon immediate decrease of the strain amplitude, $G'$ and $G''$ follow a different behavior with no peak in $G'$ and only a weak peak in $G''$, a behavior more conventionally seen in several yield stress fluids: colloidal gels and other systems such as soft colloidal glasses and granular suspensions. Note however, that the linear elastic modulus after the DSS upward – downward cycle, $G'_{DSS}$, is higher than the value at the start of the DSS cycle, $G'_{Rej}$. Therefore, we observe not only a transient strain hardening during the upward DSS (related to the $G'$ peak) but also a permanent enhancement of the elasticity at the end of the DSS (up-down) cycle. In Fig. 5(A) and 5(B), respectively, we report the linear elastic modulus and the yield strain at the cross-over of $G'- G''$ from the upward DSS (denoted as $G'_{Rej}$ and $\gamma_{y\ Rej}$ respectively), scaled by the corresponding values from the downward DSS (denoted $G'_{DSS}$ and $\gamma_{yDSS}$ respectively) curve, as a function of concentration for all clay dispersions studied here. In Fig. 5(A) we see a common behavior among all samples where an initial hardening regime (with a permanent strengthening of the linear elasticity after the DSS cycle) is followed by a strong softening as concentration increases. More specifically, immediately after the hardening regime, for all the samples the ratio approaches a plateau ($c \sim 1 - 4\ wt.\%$ for Pangel B5 dispersions and Mixtures) and ($c \sim 4 - 7\ wt.\%$ for VG-69 dispersions) and then at higher concentrations there is a further step down. We suspect that this occurs due to a change of the yielding process, where at lower concentrations (see Fig. 4) the suspensions show a fluid-like response with $G' \sim \gamma^{-2}$ and $G'' \sim \gamma^{-1}$, almost the double value in the exponent of $G'$ compare with that of $G''$. At higher clay concentrations, however, such a fluid-like response is not observed immediately after the crossover (see Fig. S3). Instead, over a broad range of strains ($\gamma \cong 1 - 100\%$), $G'$ and $G''$ drop with a slope $G', G'' \sim \gamma^{-1.5}$, indicating that some residual structure remains under flow and suggesting different yielding mechanisms [48]. During the upward-downward DSS we observe the gel – sol - gel transition with the DSS treated gels exhibiting clearly different elasticity and yield strain compared to the initial (rejuvenated) sample. Though the strain hardening behavior is observed in a very narrow concentration regime (just above the critical concentration of the sol – gel transition, see Fig. 3 solid lines) there are strong features which lead us to believe that hardening is an inherent property of the suspensions and not an artifact. We find that the behavior is (i) reproducible, (ii) frequency independent (0.1 to 8 rad/s, see Fig. S4 (A)-(C)) and most importantly is even more apparent where we follow a strict shear history protocol (see next section). The DSS treated dispersions can thus become 150 times weaker at high concentrations and about 4 times stronger at low concentrations. In Fig. 5(B), we can also distinguish two different yielding gel responses as determined by the yield strain represented by the cross-over strain amplitude. At lower concentrations ($c < 1\ wt.\%$) the dispersions show a weak decrease of the yield strain during the downward DSS (as compared to that in the upward one) suggesting weak to no hysteresis and with the ratio $\frac{G'_{DSS}}{G'_{Rej}}$ being almost 1. For higher clay concentrations there is a strong hysteresis with $\gamma_{y\ DSS}$ of the upward DSS taking place at significant higher strain amplitudes compared to that seen during the downward DSS. Note that some system dependent differences exist. The yield strain (see Fig. 5(B)) of the Pangel B5 dispersions drop significantly with concentration whereas that of the VG-69 dispersions is almost $c$ independent. Existing models of Aksay and Morbidelli [23,60] suggests a weaker connection of inter-flocs in the later. Regarding the mixtures, there is a transition regime, where at lower clay concentrations mixtures behave like Pangel B5 dispersions and at higher concentrations like VG-69 dispersions, as seen both in the linear response (Fig. 3) as well as the yield strain (Fig. 5(B)). The findings are independent of how we define the yield strain, as shown in Fig. S6. Mixtures behave like Pangel B5 dispersions (VG-69 dispersions) at lower (higher) concentrations while the underlying microstructural origin of such behavior will be the subject of future studies.

Scattering data (SALS, SAXS) show limited effect of shear on scattering patterns. They exclude the possibility of shear induced alignment. Independently of the preparation/shear history, (i.e., rejuvenated, DSS treated or steady shear treated) and of the clay concentration isotropic scattering patterns were always observed at rest (see Fig. S5) implying isotropic microstructure at rest.

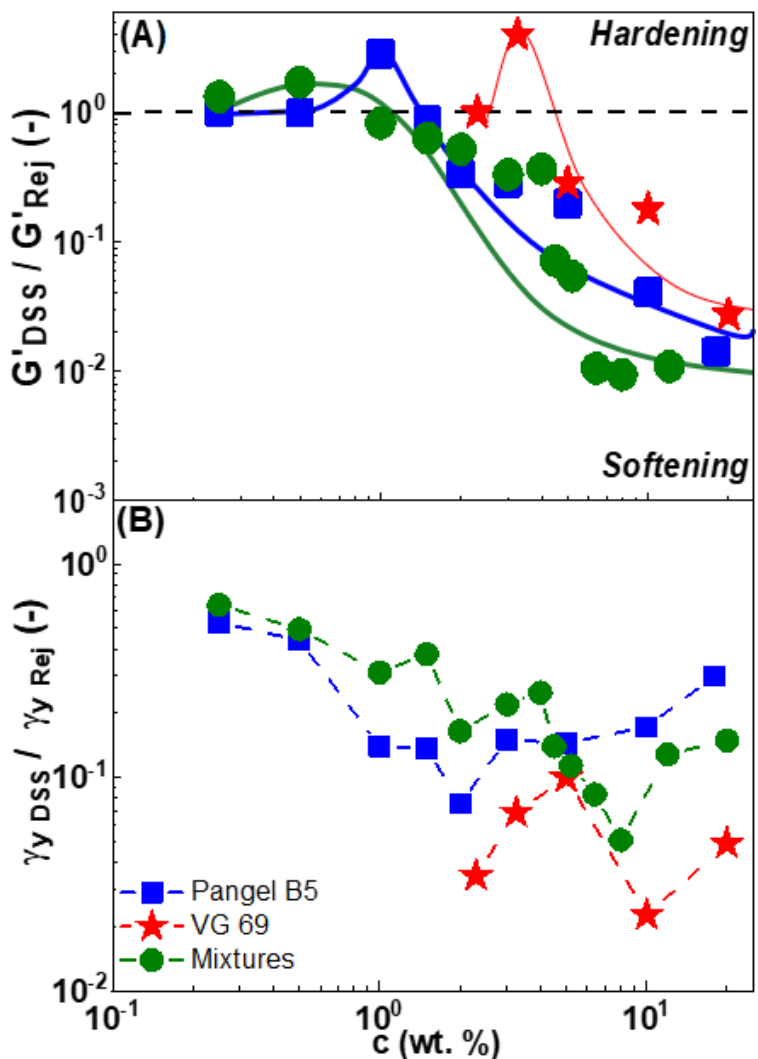

**Fig. 5:** DSS treated (A) plateau modulus ($G'_{DSS}$) and (B) crossover of $G'$ with $G''$ (yield strain, $\gamma_{y\,DSS}$) as a function of concentration of Pangel B5 dispersion (blue squares), VG-69 dispersion (red stars) and their mixtures ($50/50\,wt.\%$) (green circles). Data are rescaled with the respective plateau modulus ($G'_{Rej}$) and yield strain ($\gamma_{y\,Rej}$) value of the rejuvenated dispersions.

**D. Steady shear response at different clay concentrations**

An alternative deformation to the oscillatory DSS sweep described above is a steady shear experiment consisting of a ramp up and subsequent ramp down (or the reverse) steady shear rate sweep, or flow curve, experiment. In Figs. 6(A) and 6(B) we report such steady shear rate sweeps for different concentrations of Pangel B5 dispersions (squares) and VG-69 dispersions (stars), performed from high to low (filled symbols) and low to high (open symbols) shear rates (with direction indicated by the arrows). For all concentrations studied, the dispersions demonstrate typical yield stress response captured well by a Hershel – Bulkley (H-B) and T-C model described by Eqs. (1) and (2) and depicted with dashed lines in Figs. 6(A) and 6(B), respectively [70]:

$$\sigma = \sigma_y + k\dot{\gamma}^n \tag{1}$$

$$\sigma = \sigma_y + \sigma_y \left(\frac{\dot{\gamma}}{\dot{\gamma}_c}\right)^{0.5} + \eta_{bg}\dot{\gamma} \tag{2}$$

TC (Eq. 2) is a three-component model where the stress dissipation is described by a combination of elastic, plastic and viscous dissipation. In contrast with the famous and widely used H-B model, where k and n are arbitrary parameters without physical basis, the T-C model accounts for plastic dissipation through the critical shear rate value ($\dot{\gamma}_c$) and viscous dissipation through the $\eta_{bg}$ which in principle is the viscosity of the medium. Both models account for elastic dissipation through the yield stress value ($\sigma_y$) value. We find that yield stress, $\sigma_y$, increases sharply with clay concentration for both the single clay dispersions (see Fig. 6(C)) following $\sigma_y \sim c^{2.9}$ and a rather weaker dependence ($\sigma_y \sim c^{2.4}$) in the case of mixtures, similar to the elastic modulus shown in Fig. 3. Here, we note that in the case of mixtures, for concentrations higher than 3 wt. % a drop of the stress at lower shear rates was observed ($< 1\ s^{-1}$) probably due to slip effects or non-uniform flow. We ignored this drop for the calculation of the yield stress. The yield stress $\sigma_y$ value extracted from each of the different models is in perfect agreement, The value of the exponent *n* (H-B fit) for the shear rate dependence at high shear rates is between 0.5 (for high) and ~ 0.9 (lower concentrations) indicating a sublinear to linear shear thinning behavior depending on the sample and concentration and depicted in Fig. 6 (D). In Fig. 6 (E) we report on the evolution of the critical shear rate, ($\dot{\gamma}_c$) (T-C fit). Most importantly the $\dot{\gamma}_c$ value increases monotonically (n from H-B decreases monotonically) with clay concentration which implies important changes of plastic and viscous dissipation. Performing rheo – SAXS and rheo-SALS measurements under similar steady shear rates in a Pangel B5 dispersion we find a weak shear-induced anisotropy (along the shear-vorticity plane) that disappears immediately (in order of a second) after flow cessation, implying that the needles re-orient isotropically after flow cessation (see Fig. S7). This is in agreement with previous experiments with clay - cellulose nanocrystal mixtures where it was found that the addition of the latter delays significantly the orientation of particles after flow cessation [71,72]. Also note that, for all samples, under the current measurement conditions the hysteresis between up-down steady shear sweeps is quite weak, significantly less apparent compared to large thixotropic effects in soft glassy systems or water based clay gels [73,74].

Hysteresis is also significantly weaker in flow curve than in the LAOS measurements in Fig.4. The determination of the timescales involved in thixotropic effects (and their comparison with timescales related with the more permanent effects involved in shear induced tuning of mechanical properties) should be the subject of further studies [20,21,75,76]. The difference between the flow curve and the LAOS conditions and mostly the different duration of the tests (~3 min for flow curves, ~60min for LAOS) provide a indication that it should be in the range of minutes.

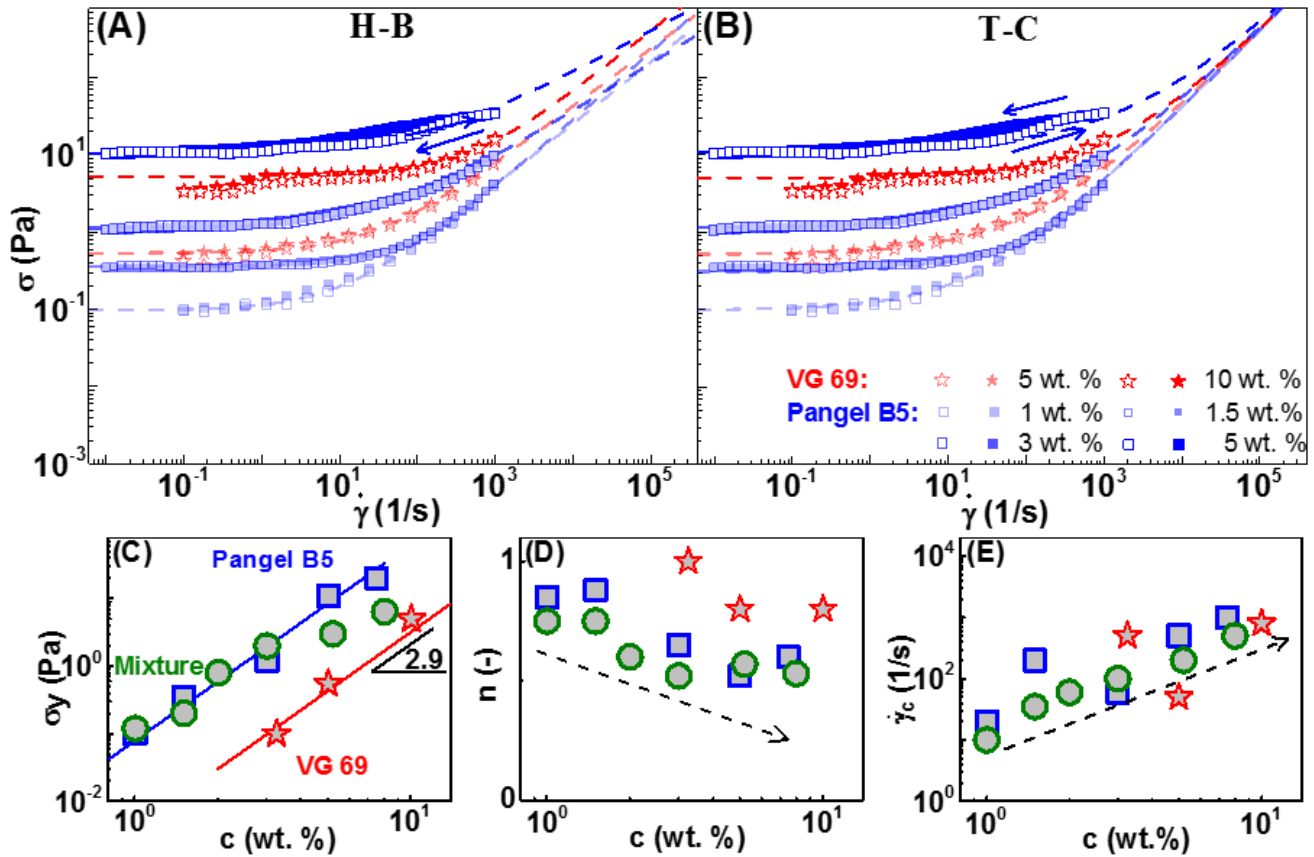

**Fig. 6:** (A), (B) Flow curves at different concentrations of VG-69 dispersion (red stars) and Pangel B5 dispersion (blue squares). Measurements performed from high to low (filled symbols) and low to high (open symbols) shear rates. Dashed lines in (A) represent H-B and in (B) T-C fits (see ref. [70]) of the experimental data. (C) yield stress, $\sigma_y$ (D) exponent $n$ from H-B model and (E) critical shear rate ($\dot{\gamma}_c$) as a function of concentration for Pangel B5 dispersions (blue squares), VG-69 dispersions (red stars) and mixtures of them (green circles).

**E. Tuning of the elasticity of clay dispersions: Steady versus oscillatory preshear protocols**

Shear history affects the linear and non-linear rheological properties as well as the microstructure of colloidal gels [3,49,50]. As the thixotropic effects during the DSS and flow curve cycles are indicative of a strong preshear history dependence we have followed a specific preshear protocol to probe in a robust and reproducible way such effects. As shown schematically in Fig. 7, we first performed high shear ($\dot{\gamma}$ = $1000 s^{-1}$) for $60\ s$ which constitutes rejuvenation that erases all prior shear history of the sample. Then we performed either steady shear flow at a constant shear rate ($0.01 < \dot{\gamma} < 1000 s^{-1}$) or large amplitude oscillatory shear (LAOS) measurement ($\gamma \geq 2\%\ at\ \omega = 1\ rad\ s^{-1}$) for $200\ s$. These tests constitute our preshear preparation protocol. The evolution of the viscoelastic properties after the preshear sequence was probed by small amplitude dynamic time sweep (DTS) for $1200\ s$ and a subsequent dynamic frequency sweep (DFS). Both DTS and DFS were performed in the linear regime with strain amplitudes ranging from $\gamma = 0.1$ to $1\%$, depending on the sample.

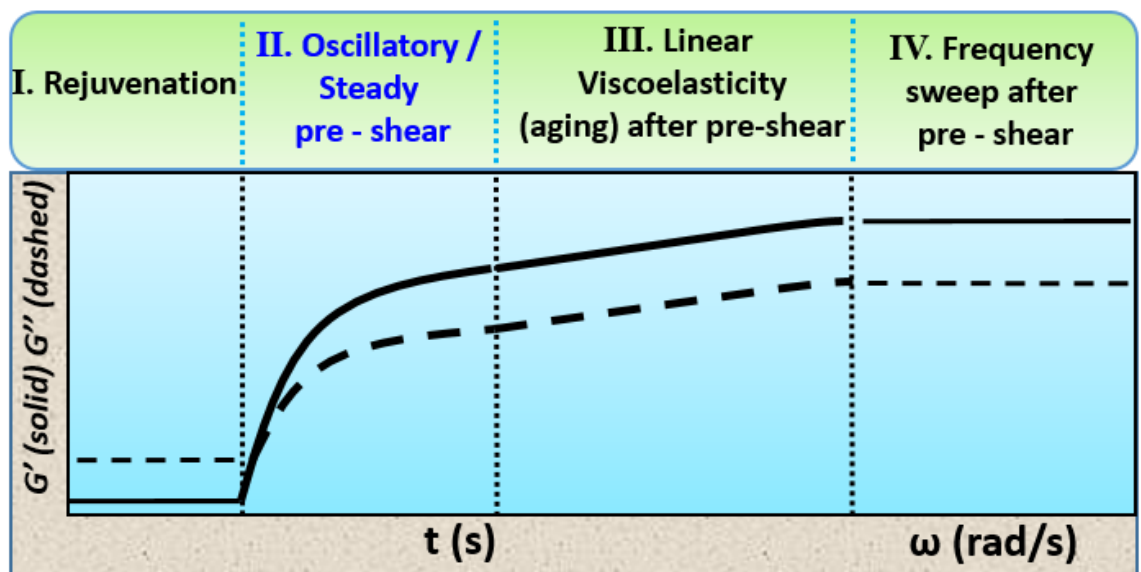

**Fig. 7:** Rheological protocol followed for the shear induced tuning of the clay dispersions. $G'$ and $G''$ are indicatively shown for (I) rejuvenation consisting of high shear rate ($\dot{\gamma} = 1000 s^{-1}$), (II) steady or oscillatory preshear, (III) probing of the linear viscoelasticity at specific frequency over time with a DTS and (IV) a linear viscoelastic spectrum measured by a DFS.

In Fig. 8 we report the evolution of $G'$ (filled symbols) and $G''$ (open symbols) during the tests of the above shear protocol for a Pangel B5 dispersion ($c = 1\ wt.\%$). As can be seen in Fig. 8 (A), in regimes II and III, the application of medium amplitude oscillatory preshear ($\gamma = 15\%$, as indicated by blue lines) induces increased elastic modulus of the dispersion, as compared to the rejuvenated sample elasticity (represented by the $\gamma = 0.1\ \%$ black curves in Figs. 8(A) and 8(B)). Similar data are shown in SI, Figs. S8 and S9. It should be noted that the increased elasticity (or strain hardening) by the application of certain strain amplitude oscillatory preshear is detected in all dispersions at lower concentrations in agreement with the observed behavior during the DSS treatment discussed above (section III.C).

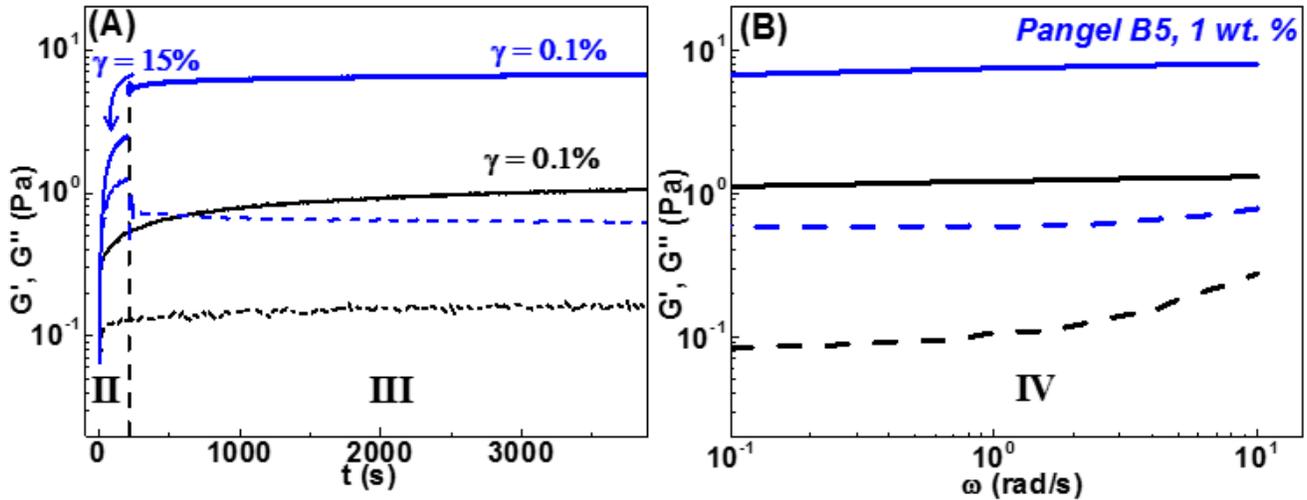

**Fig. 8:** Storage $G'$ (solid lines) and loss modulus, $G''$ (dashed lines) for a Pangel B5 dispersion (1 $wt.\%$) at different regimes during the shear-tuning protocol as evolved (A) with time in period II ($\gamma = 0.1\%$ for black and $\gamma = 15\%$ for blue lines) and III (after flow cessation with blue lines stepping from $\gamma = 15\%$ to $\gamma = 0.1\%$) DTS at $\omega = 1\ rad/s$ and (B) with frequency (DFS at $\gamma = 0.1\%$) in period IV.

We next (Fig. 9) directly compare the effects of steady (blue) and oscillatory (grey - light cyan symbols) preshear tuning at two different Pangel B5 dispersion concentrations (1.5 $and$ 5 $wt\ \%$). In Fig. 9 we gather the elastic modulus values (at a specific frequency) after preshear for different clay gels. For clarity, we normalize the values of $G'$ extracted after different preshear treatments with the value of $G'$ obtained after rejuvenation (denoted here as $G'_{Rej}$). The scaled elastic modulus is plotted as a function of the shear rate (x-axis) in the case of steady preshear treatment or of $\gamma_0\omega$, i.e. the maximum intracycle shear rate in the case of oscillatory preshear. We find that the results of the two protocols, (i.e. the steady and the oscillatory preshear), coincide well though the largest applicable rate during the oscillatory shear does not correspond to the high steady shear rate used during the shear rejuvenation step. We should note that this is not a trivial finding, as in other colloidal systems, such as depletion gels, oscillatory and steady preshear does not necessarily lead to same effects, with oscillatory preshear usually more efficient in tuning the elasticity of the sample. Furthermore, here we see two distinct behaviors. At the low concentration we see a shear induced strengthening (with increase of $G'$) for dispersions prepared through steady or oscillatory shear at low rates (possible increase number of contacts) whereas for the higher concentration (5 $wt.\%$) a softening is apparent in the entire shear rate regime (possible decrease number of contacts per stack), but with stronger effect at intermediate rates/strain amplitudes. Since there is a good agreement between the two preshear protocols, below we further elaborate on findings from oscillatory shear treatment of the different dispersions. Shear may have an impact on distribution of contacts between clay particles with an interplay between inter and intra floc interactions, for example by creating a more or less dense array of contacts. Contacts are notoriously difficult to identify through scattering experiments. If shear induces only very local changes in the contacts keeping the overall structure largely the same, this might not be easily observed by scattering techniques (cartoon of Fig. 9).

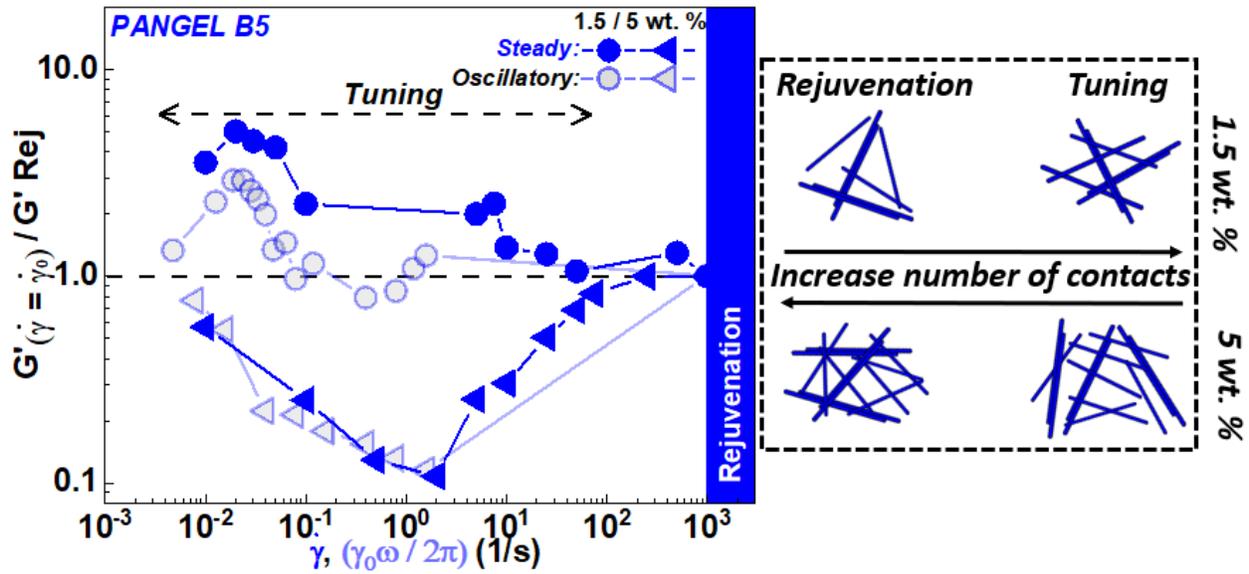

**Fig. 9:** Plateau modulus of Pangel B5 dispersions at $1.5\ wt.\%$ (circles) and $5\ wt.\%$ (triangles) prepared at different steady shear rates (filled blue) or oscillatory strains ($\omega = 1\ rad/s$) (grey-light blue symbols) prior to the linear measurement. Data is normalized with the plateau modulus of the prepared dispersion at $1000\ s^{-1}$, which is labelled $G'_{Rej}$). Cartoon in the right side shows out of scale needles packed together after rejuvenation or tuning. Arrows indicate increase of contacts between needles.

**F. Tuning of clay dispersions through the oscillatory shear - concentration dependence**

In Figs. 10(A)-10(C) we show the effect of oscillatory preshear on each clay dispersion. For the Pangel B5 dispersion (Fig. 4(A)), at lower clay concentrations ($c < 2\ wt.\%$, transparent blue symbols) we see that gels presheared at lower strain amplitudes become stronger compared to those after rejuvenation (i.e. prepared by high preshear rates). At $c = 2\ wt\%$ the preshear treatment seems to have no effect on the dispersion elasticity whereas at higher concentrations ($c > 2\ wt\%$, blue symbols) preshear leads to weakening (decrease in the elasticity) of the sample. The same trend is found for VG-69 dispersions, (Fig.

10(B)) while for the mixtures (50/50 $wt.\%$) (Fig. 10(C)) only a very weak strengthening is detected at the lowest concentration at intermediate strain amplitudes but a rather strong weakening at higher concentrations. In general, the transition from shear induced strengthening to weakening takes place at higher clay concentration in the VG-69 dispersions compared to the Pangel B5 dispersions.

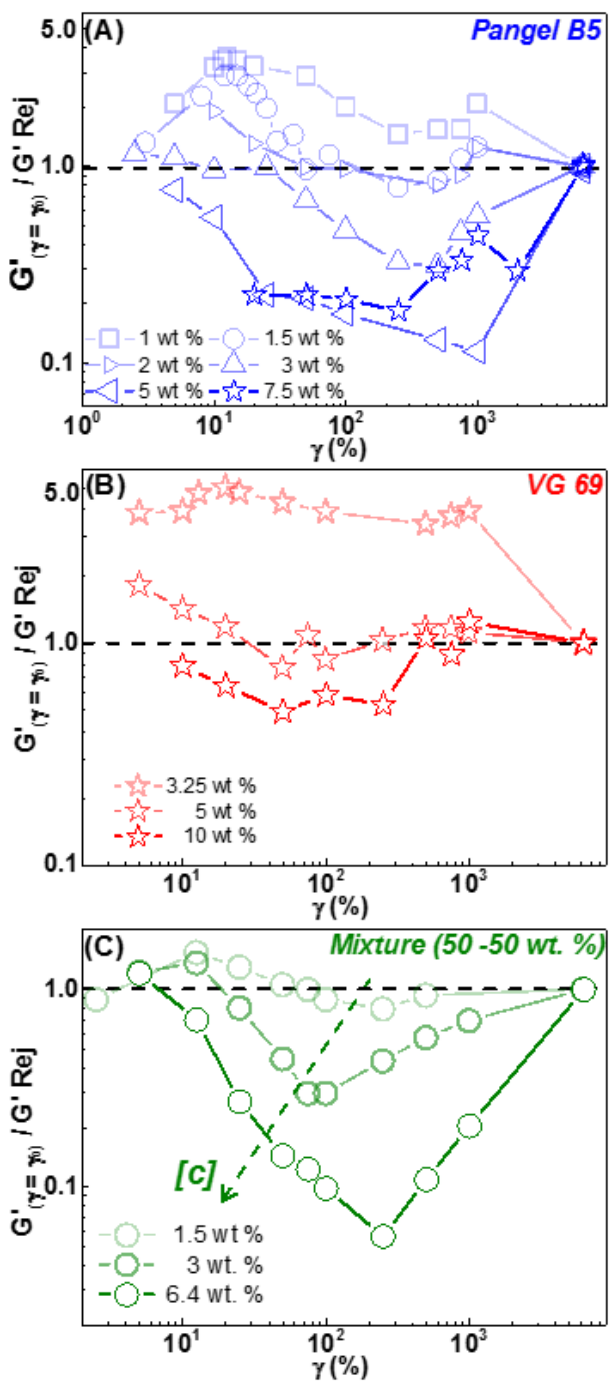

**Fig. 10:** Plateau modulus of (A) Pangel B5 dispersions, (B) VG-69 dispersions and (C) mixtures (50 − 50 $wt.\%$) prepared at different oscillatory strains prior to the linear measurement as a function of oscillatory strain amplitude. Data are normalized with the plateau modulus of the prepared dispersion at $1000\ s^{-1}$, $G'_{Rej}$). Arrows indicate increasing concentration.

The elastic modulus from these experiments as a function of clay concentration are summarized in Figs. 11(A) and 11(B). From this plot, is becomes clear that we detect a power law increase of the elasticity (Fig. 11(A)) and decrease of yield strain (Fig. 11(B)) with increasing concentration, with power law exponents that depend on the preshear treatment of the samples. The overall response is a result of the different effects preshear has on the sample elasticity at low and high concentrations (as seen in Figs. 9 and 10 and discussed above). With respect to the fully rejuvenated Pangel B5 dispersion, preshear leads to a strengthening of the sample at low concentrations and a weakening at high concentrations. Both effects are balanced at a specific intermediate strain amplitude as seen in Fig. 10, leading to a decrease of the power law exponent of the dependency of $G'$ on concentration from a value of 3.9 (for the fully rejuvenated sample) to as low as 2.5 for a presheared sample. Moreover, the yield stress of the samples evolved with shear-treatment. Assuming the yield stress from yield strain and elastic modulus ($\sigma_y \sim G_P\ \gamma_y$), we saw that it depends both on shear treatment and on clay concentration (see Fig. S11). Similar changes in the power law increase of $G'$ vs. concentration, depending on the preshear conditions is also seen in the VG-69 dispersions (Fig. 11(A)). Although changes in the slope of the elasticity with concentration depending on the shear history were reported previously [41–43] such spectacular behavior with a universal weakening (or softening) and strengthening (or hardening) regimes is to the best of our knowledge reported for the first time. The changes of the scaling laws should also reflect changes in microstructure and the corresponding fractal dimension of the sample. Most existing models relate the concentration dependence of the plateau elastic modulus, $G_P$, with the fractal dimension as measured by scattering experiments. Therefore, change in the $G_P$ versus concentration scaling should mean change of fractal dimension. Here however this seems not to be the case as Rheo-SAXS measurements mentioned above did not detect any clear changes in the scattering intensity, $I$, versus the scattering wave-vector, $q$ [60,61]. Hence, we detect a qualitative change in linear viscoelasticity that is not accompanied by a qualitative change in the microstructure, at least in the length-scales probed by our scattering experiments. One reason could be that the rheology is controlled by changes in the bond connectivity (or more generally structural changes), for which SAXS is not sensitive. For example, a relevant parameter which cannot be assessed through macroscopic rheological measurement is the number of bonds contributing to the gel elasticity (in other words bond connectivity) and how this varies on the preshear history/treatment [77–80].

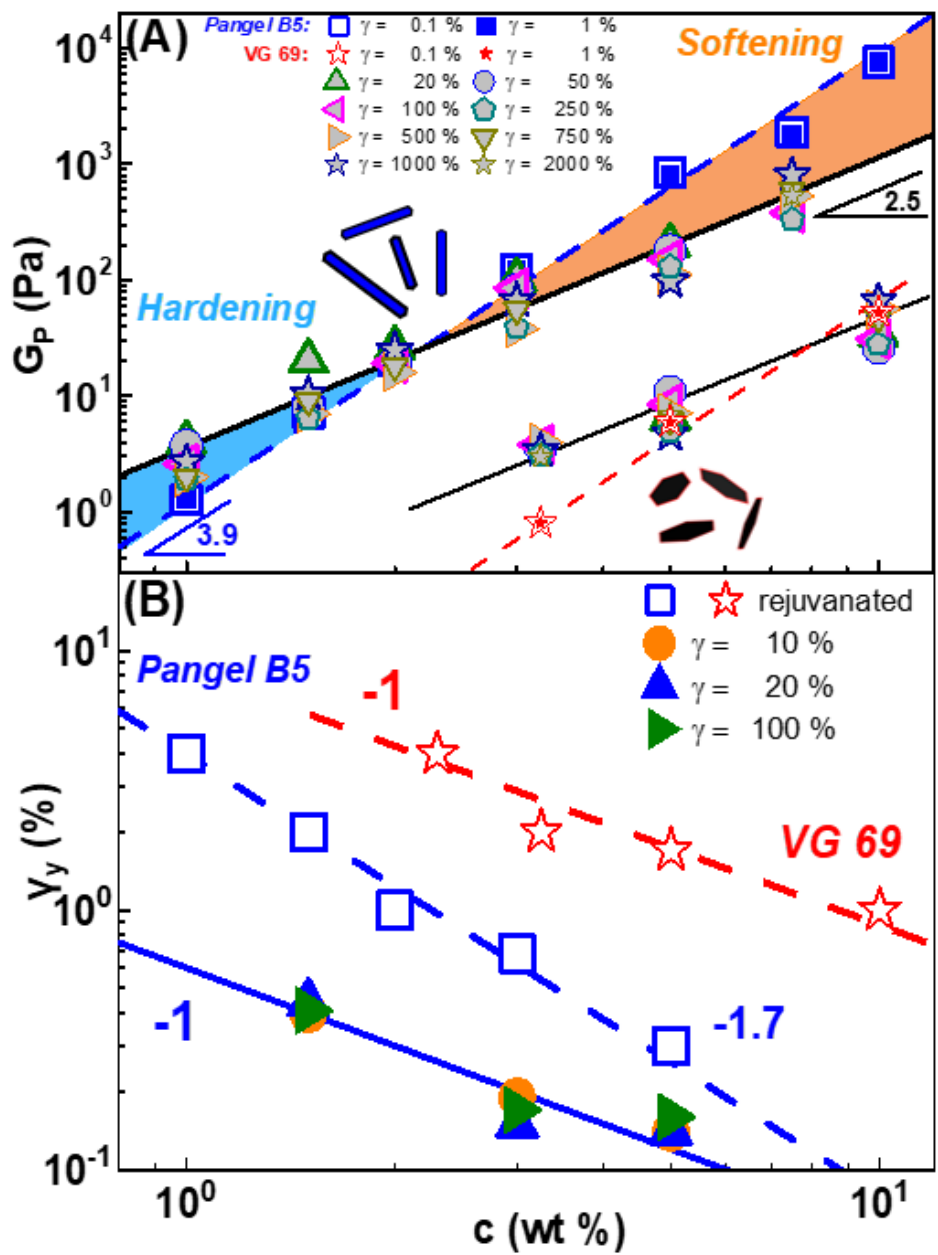

**Fig. 11:** (A) Plateau modulus ($G_P$) as a function of clay concentration of the needle-like Pangel B5 and plate-like VG-69 colloidal dispersions presheared at different strain amplitudes prior to the linear measurement ($\omega = 1\ rad/s, \gamma = 0.1\%$). Blue squares correspond to the shear rejuvenated Pangel B5 dispersions whereas red stars to the shear rejuvenated VG-69 dispersions. (B) Corresponding yield strain values (as deduced from the deviation of G'' from linearity) for needle-like Pangel B5 samples (derived from Fig. S10) and plate-like VG-69 samples (derived from Fig. S6(A)).

## IV. Discussion

The relation between structure and elasticity and its dependency on the shear history of colloidal gels has long been discussed and is still subjects of many studies [81–83] aiming to relate rheological findings on the gel elasticity and yield strain with underlying changes in the gel microstructure. The power law concentration dependence of plateau modulus $G_P \sim \varphi^A$ and yield strain (as deduced from the deviation of G'' from linearity) $\gamma_y \sim \varphi^B$ are often discussed in the frame of fractal gels models where the power laws exponents A and B are expressed in term of the fractal dimension of the gel. We here attempt such an analysis though the fractal nature of organoclay gels has not been verified. Fractal gels models assume that the networks comprise close packed fractal flocs (described by a backbone fractal dimension $x$, with values typically between 1 and 1.3) forming gels of fractal dimension, $d_f$ [23,60,61]. The work of Shih et.al. [23] considers two extreme regimes, the weak and the strong link limit, where the mechanical response of the aggregated flocs is dominated by the intra or the inter-floc links respectively. In the weak link regime, the power law exponents describing the volume fraction dependence of the elastic modulus and yield strain are equal. With $A = B = \frac{1}{(d-d_f)}$ (with $d$ the space dimension); therefore both $G_P$ and $\gamma_y$ increase with concentration. On the other hand in the strong-link regime $A = \frac{(3+x)}{(3-d_f)}$ and $B = \frac{-(1+x)}{(3-d_f)}$; therefore $G_P$ increases whereas $\gamma_y$ decreases with concentration (increase of fragility) [23].

In our clay dispersions, we get a concentration dependence $G_P \sim c^{3.9}$ both for VG-69 and Pangel B5 dispersions but the dependence of the yield strain differs slightly for the two materials with $\gamma_y \sim c^{-1}$ and $\gamma_y \sim c^{-1.7}$ respectively. These values are more in line with the values reported in the strong link regime, as the decreasing yield strain excludes the predictions of the weak link limit. However, using A and B to evaluate the fractal dimensions $d_f$ and $x$ returns reasonable value of $d_f$, but unrealistic (less than unit or even negative) values of the backbone fractal dimension $x$ as can be seen in table S1.

Morbidelli et. al. [60] proposed an extension of the original model to cover the intermediate regime between the strong and weak link limits, introducing a constant $a$ representing the ratio of intra floc and inter floc elasticity that allows to account for cases where the strength of intra and inter floc links are comparable and therefore their contribution to the overall gel elasticity cannot be decomposed . In this model the exponents are expressed as: $A = \frac{\beta}{(3-d_f)}$, $B = \frac{(3-\beta-1)}{(3-d_f)}$, with, $\beta = 1 + (2+x)(1-a)$.

The two extreme cases are recovered as the weak – link regime with $\alpha = 1$, so, $\beta = 1$, and the strong link regime with $\alpha = 0$ and $\beta = 3 + x$. The use of the transition model [60] in our data leads to reasonable values of $d_f$ and $x$ for all dispersions studied here (table S1).

In particular it suggests that the difference in the exponents of plateau modulus (2.5 to 3.9, Fig. 11(A)) and the yield strain ($-1$ to $-1.7$, Fig. 11(B)) for the pre-shear treated and the rejuvenated Pangel B5 could originate from a change in the ratio of the inter- to intra-floc contribution to the modulus. Therefore, the application of this model offers a possible microscopic origin for the changes of the macroscopic mechanical response observed upon rejuvenation i.e. that the latter enhances the inter-cluster elastic contributions (where $\alpha$ approaches 0).

The deduced $d_f = 2.16$ value for the mixture ($50/50\ wt.\%$) lies between those of the Pangel B5 and the VG-69 whereas $a \sim 0.5$ is close to those of VG-69 samples rather than the Pangel B5. According to the model, intra-cluster elastic contributions are more important in VG-69 and the mixtures compared to Pangel B5.

Though the values of the transition model seem reasonable; we note again that the application of a fractal model for organoclay particles with high aspect ratio may be questioned. It should be also noted that the fractal dimension of the rejuvenated Pangel B5 dispersion extracted from rheology ($d_f = 2.09$) is in disparity with the value deduced from scattering ($I \sim q^{-2.5}$, depicted in Fig.S5(C)). For the VG 69, the apparent fractal dimension from rheology ($d_f = 2.31$) and scattering ($I \sim q^{-2.5}$, [58]) are closer.

However, we should point out that in all cases (both for the VG-69 and the Pangel B5), the use of the scattering fractal exponent, $d_f$, in the Morbidelli model leads to equations for A and B without a common solution for x and $\alpha$.

Moreover the structural studies of organo-clay platelet systems [57,58] suggest a formation of large platelet aggregates (referred to as tactoids) for which the apparent power laws from scattering and rheology do not reflect the gel fractal dimension but rather an intra floc (tactoid) structure, and therefore cannot be used as a proof of a fractal type aggregation. Computer simulation able to trace individual particles would probably be a good approach to resolve some of the issues concerning the relationship between the structure and the rheological properties [84].

## V. Conclusions

We investigated gels of two differently shaped organoclay particles and their mixtures and we report how preshear treatment can be used to tune their final rheological properties. For the individual clay dispersions, comparing the VG-69 platelets and Pangel B5 needles, we find an identical scaling law behavior of the elastic modulus with concentration ($G_P \sim c^{3.9}$) but a slightly different yield strain dependence with $\gamma_y \sim c^{-1}$ for the VG-69 and $\gamma_y \sim c^{-1.7}$ for the Pangel B5.

In the clay mixture we observe a behavior where the plateau modulus, $G_P$, follows one of the two individual components' response, depending on the concentration. At low concentrations the mixture behaves like the corresponding Pangel B5 sample whereas at higher concentration like the corresponding VG-69 sample. As a result, we observe an overall weaker scaling-law behavior of the plateau modulus ($G_P \sim c^{2.8}$) and the yield strain ($\gamma_y \sim c^{-0.4}$). This behavior is clear evidence that there is a strong coupling between the two networks of plates and rods which affect the balance between inter and intra-floc connections and overall structure. A tentative sketch of structure of mixture (along [67]), is shown in Fig. S12, at low and high clay concentration.

Furthermore, dynamic strain sweep experiments reveal a very intriguing strong hysteresis and softening (up to 150 times) of the gel at high concentration for all systems studied. This motivated us to study shear history effects, for which we followed a strict preshear protocol. We find that upon cessation of preshear flow the elasticity may vary significantly, hence preshear can lead to gels either 5 times stronger or 10 times weaker depending on the concentration regime. Pangel B5 dispersions prepared at different shear strain amplitudes have significantly different concentration scaling law behaviors (compared with the fully rejuvenated sample), with $G_P \sim c^{2.5}$, and $\gamma_y \sim c^{-1}$, though they yield at significantly lower strain amplitudes. It implies a large change of the yield stress ($\sigma_y \sim G_P \gamma_y$) concentration scaling between shear-treated ($\sigma_y \sim c^{1.5}$) and rejuvenated dispersions ($\sigma_y \sim c^{2.2}$). Since our clay gels exhibit rheological properties and structures are reminiscent of fractal gels, the application of existing theoretical elasticity models to extract a fractal dimension indicate a combined contribution of inter and intra-floc links in the elasticity, thus belonging in a transition regime rather than in the weak or strong link limit. Thus, extracting a robust value of network fractal dimensions becomes more challenging with model fitting involving a few free parameters. Rheo-SAXS experiments acquired along the flow curve measurement suggest the existence of only a weak anisotropy under flow and relaxation to fully isotropic state immediately after flow cessation with no measurable changes in the scattering intensity. It is thus clear that preshear treatment can significantly affect the rheology without much change in the clay arrangement within the dispersions. The present findings should have a wide range of applications in design and tuning of industrial systems such as drilling fluids, or others, in a cost-effective way.

## Supplementary Material

The supplementary material online contains additional experimental rheology and scattering results and tentative cartoons of mixture clay dispersions.

## Acknowledgments

The authors would like to thank T. Gibaud, W. Chèvremont and T. Narayanan for their help with the Rheo – SAXS measurements and the fruitful discussions. ESRF is acknowledged for the provision of synchrotron beamtime (proposal number IH-SC-/785). The authors would also like to especially thank S. Papadakis at the Electron Microscopy Unit "Vassilis Galanopoulos" of the University of Crete for assistance with SEM imaging. Partial support has been received by the Twinning project FORGREENSOFT (Number: 101078989 under HORIZON WIDERA-2021-ACCESS-03). SLB is thanked for permission to publish this work.

# Supporting Information

## Tuning the mechanical properties of organophilic clay dispersions: particle composition and preshear history effects

*Nikolaos A. Burger[1,2], Benoit Loppinet[1], Andrew Clarke[3] and George Petekidis[1,2]*

[1] *IESL-FORTH, P.O. Box 1527, GR-711 10 Heraklion, Greece*
[2] *Department of Materials Science & Technology, University of Crete, Heraklion 70013, Greece*
[3] *SLB, Schlumberger Cambridge Research Ltd., High Cross, Madingley Road, Cambridge CB3 0EL, UK*

***Section I:*** *Effect of preshear and aging at different concentrations of clay dispersions*

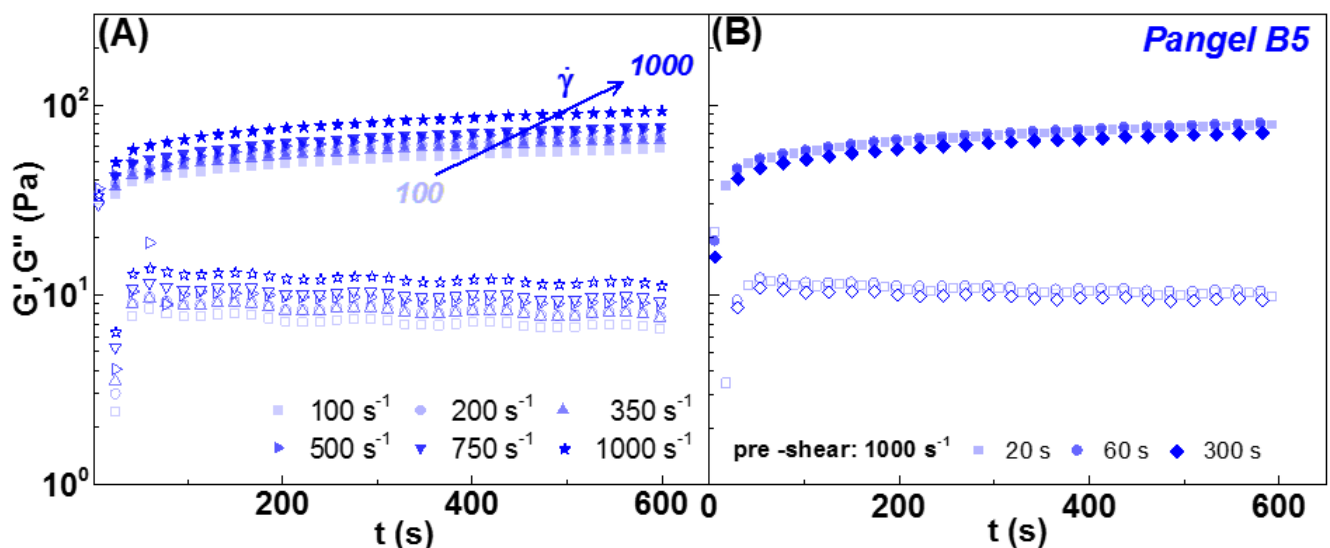

**Fig. S1:** Evolution of storage ($G'$) and loss ($G''$) modulus with time prepared at different (A) shear rates and (B) times at 1000 s$^{-1}$ for a Pangel B5 dispersion at 3 wt. %. Measurements were performed at $\omega = 1$ rad/s, $\gamma = 1\%$.

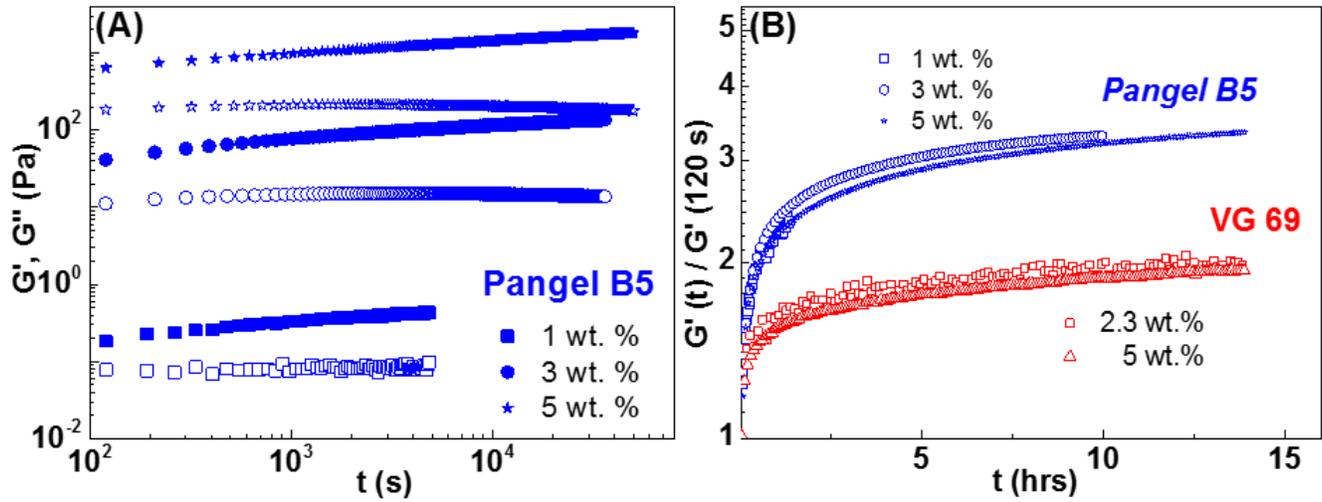

**Fig. S2:** Long – time evolution of $G'$ (filled) and $G''$ (open symbols) at: (A) different concentrations of Pangel B5 samples and (B) $G'(t)$ (rescaled with the storage modulus at $120\ s$) of Pangel B5 dispersions (blue symbols) and VG-69 dispersions (red symbols). Measurements were performed at $\omega = 1\ rad/s, \gamma = 1\%$.

## *Section II:* *Oscillatory shear of clay dispersions*

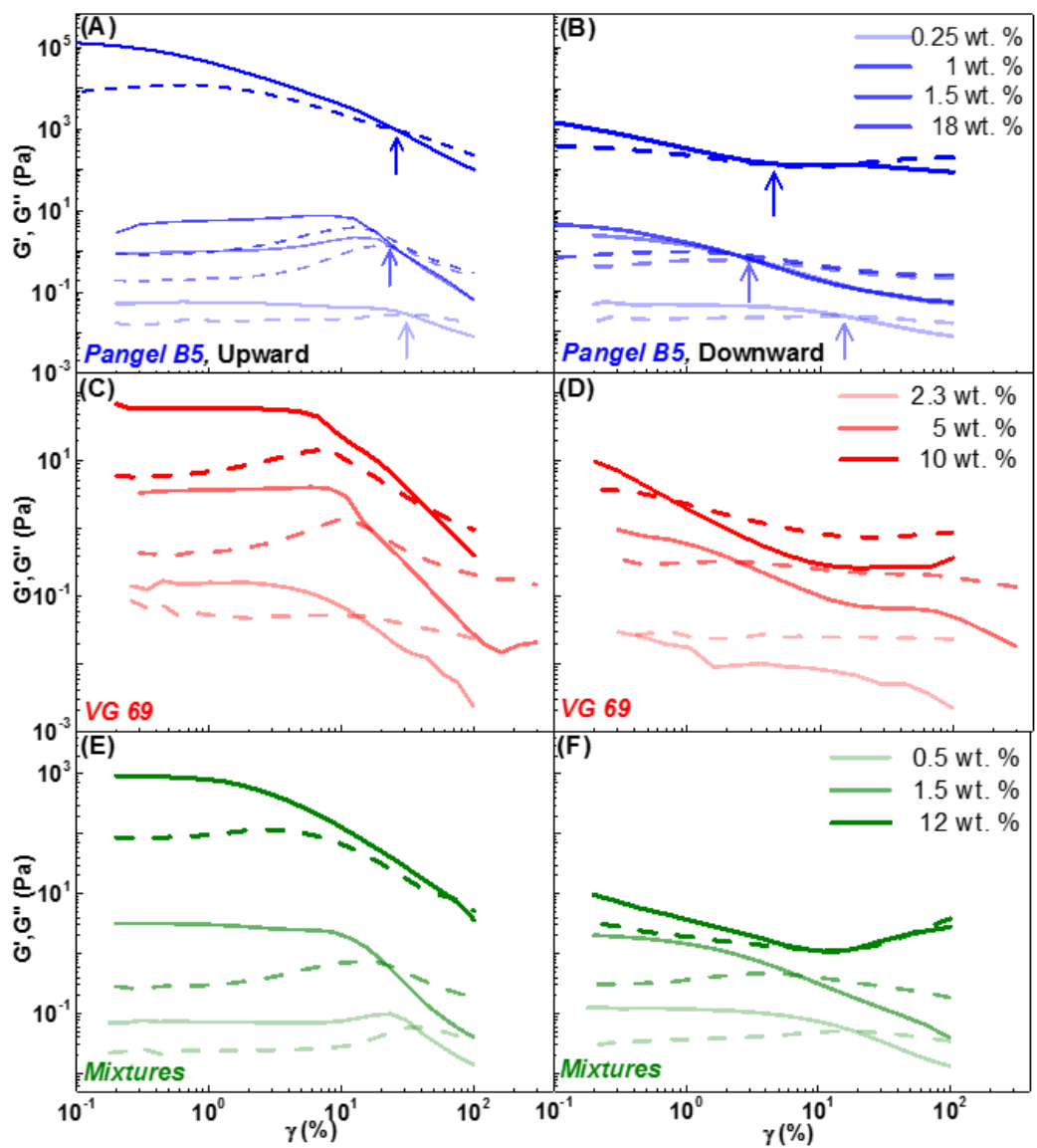

**Fig. S3:** Dynamic strain sweep measurements performed at different clay concentrations from: (A, C, E) low to high ($\gamma = 0.2 - 100\%$) and (B, D, F) high to low ($\gamma = 100 - 0.2\%$) strain amplitudes for Pangel B5 samples (A, B), VG-69 sample (C, D) and mixtures at $\omega = 1 \, rad/s$. For clarity we kept the same scaling for all the plots.

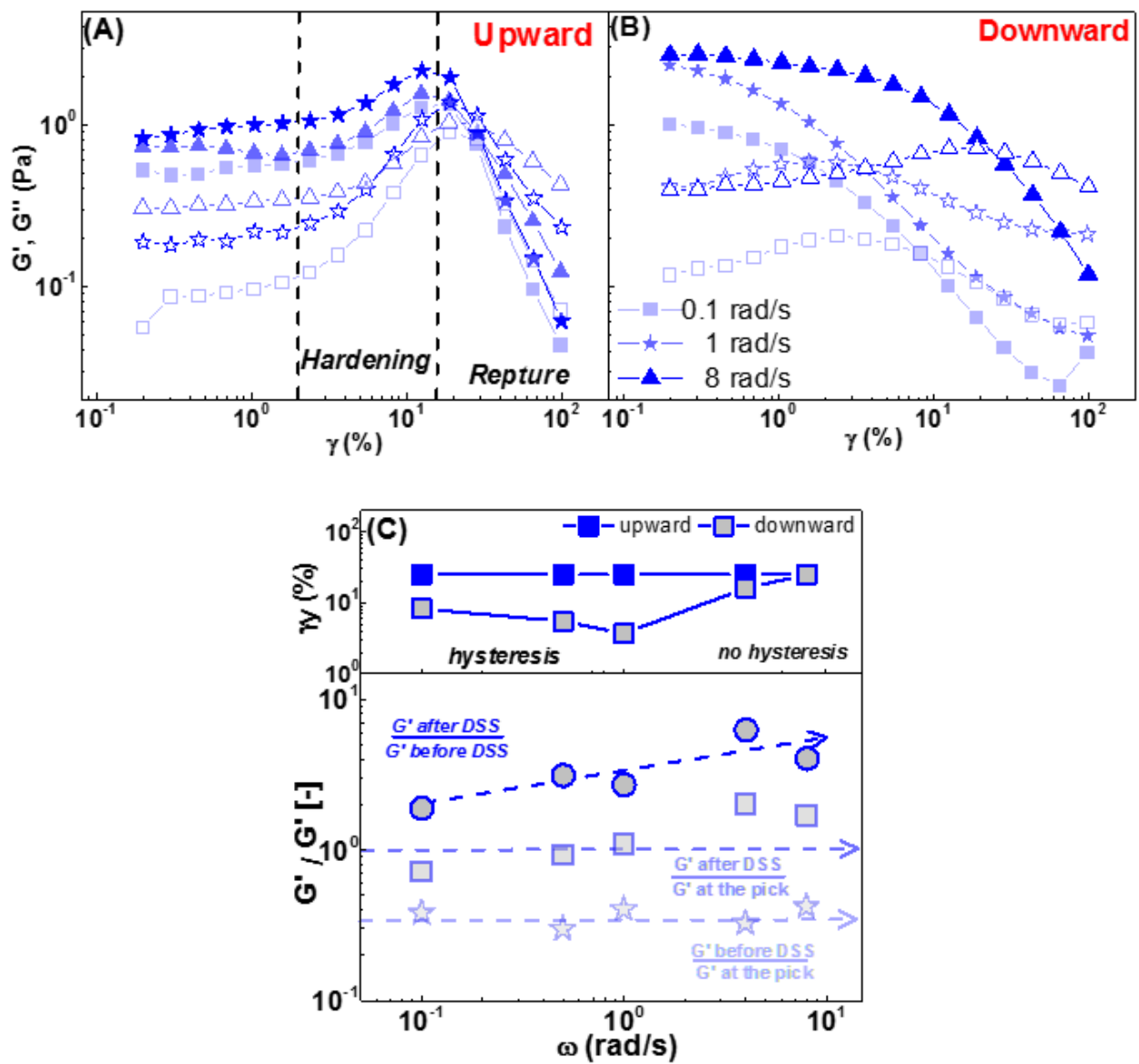

**Fig. S4:** Dynamic strain sweep measurements performed at different frequencies at $0.1 < \omega < 10 \, rad/s$ from: (A) low to high and (B) high to low strain amplitudes for a Pangel B5 dispersion at $1 \, wt.\%$. (C) Upper plot: Cross over yield strain during upward (filled) and downward (open symbols) DSS at $0.1 < \omega < 10 \, rad/s$. bottom: Ratios of plateau modulus before and after the DSS treatment as a function of frequency.

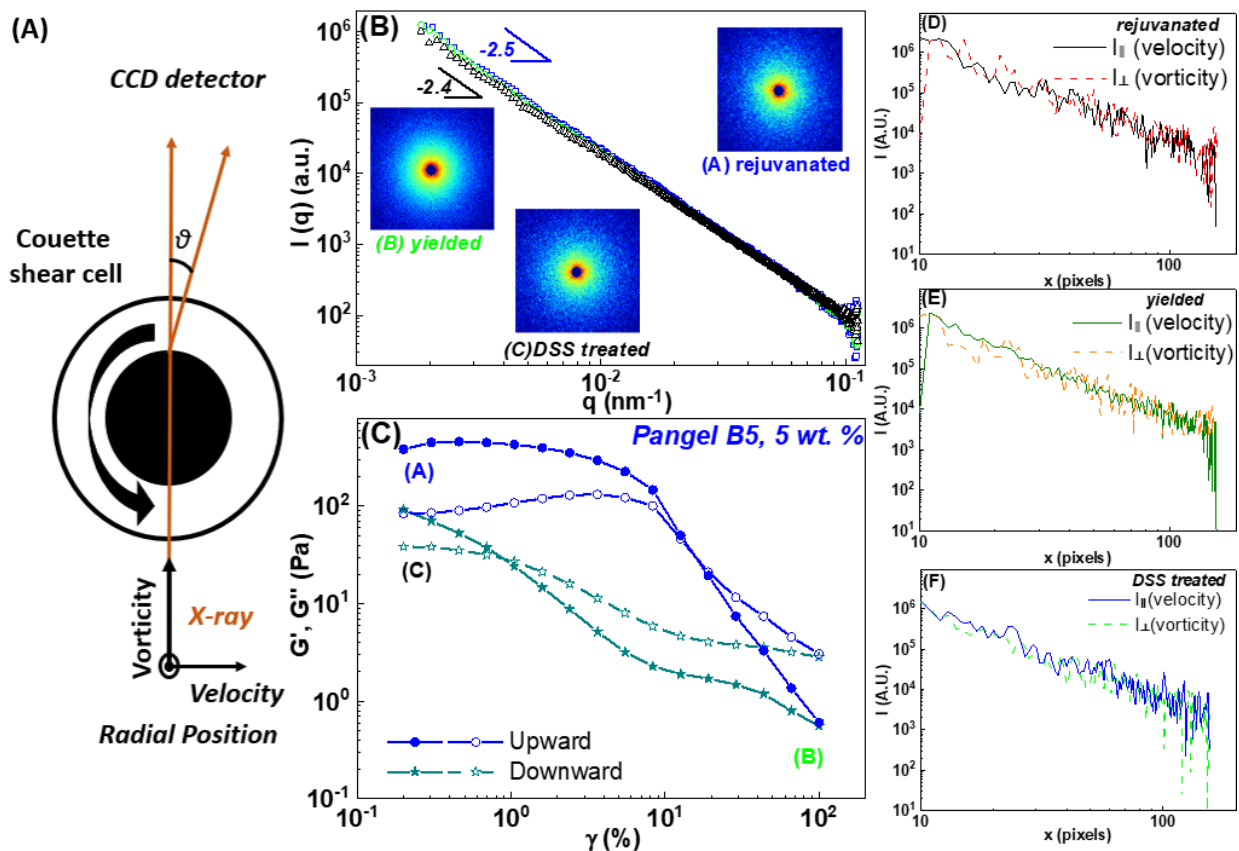

**Fig. S5:** Pangel B5 dispersion ($c = 5wt.\%$) under oscillatory shear. (A) Cartoon of rheo-SAXS geometry. illustrating the radial incidence of X-ray beam in couette cell. (B) Dynamic strain sweep measurements performed during the SAXS measurement from: low to high (blue) and high to low (dark cyan) strain amplitudes at $\omega = 1\,rad/s$. (C) Azimuthal average scattering Intensity $I$ vs scattering wave vector amplitude $q$. Inset: Corresponding scattering 2D-patterns. Intensity profiles in flow and vorticity plane at: (D) after rejuvenation, (E) yielded, (F) DSS treated.

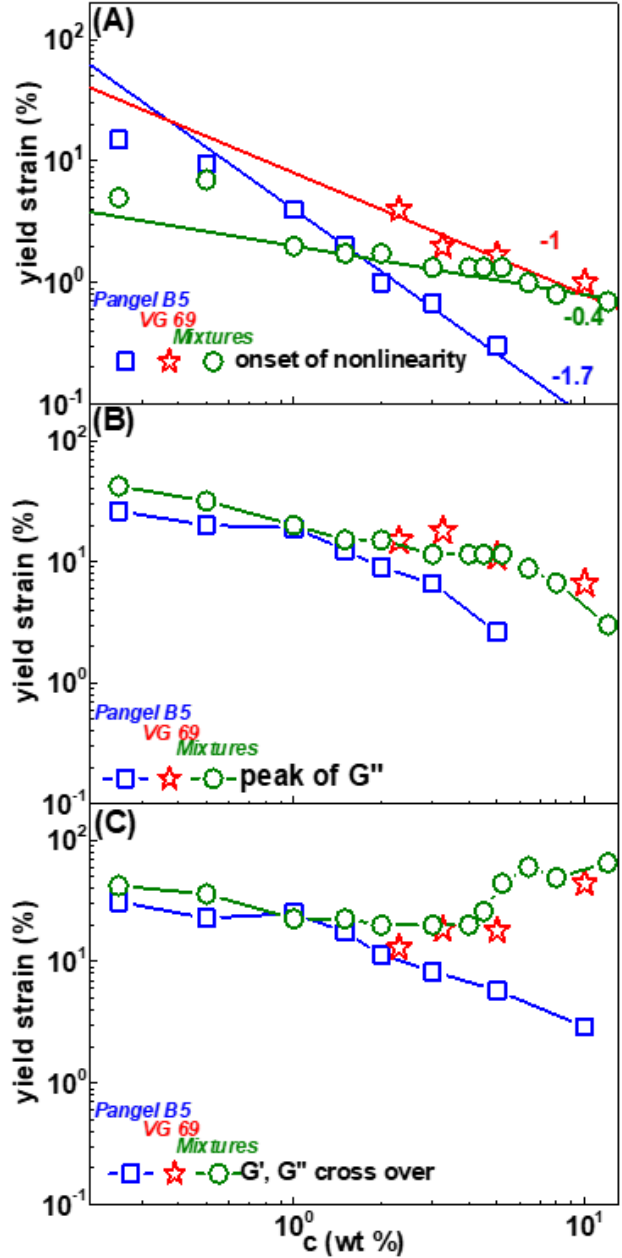

**Fig. S6:** Yield strain as a function of clay concentrations based on different definitions for Pangel B5 dispersions (blue), VG-69 dispersions (red) and mixtures (green symbols). (A) Onset of nonlinearity, (B) peak of $G''$ and (C) cross-over (yield strain) of $G'$ with $G''$. For clarity we kept the same scaling for all the plots.

*Section III:* Steady shear Measurements

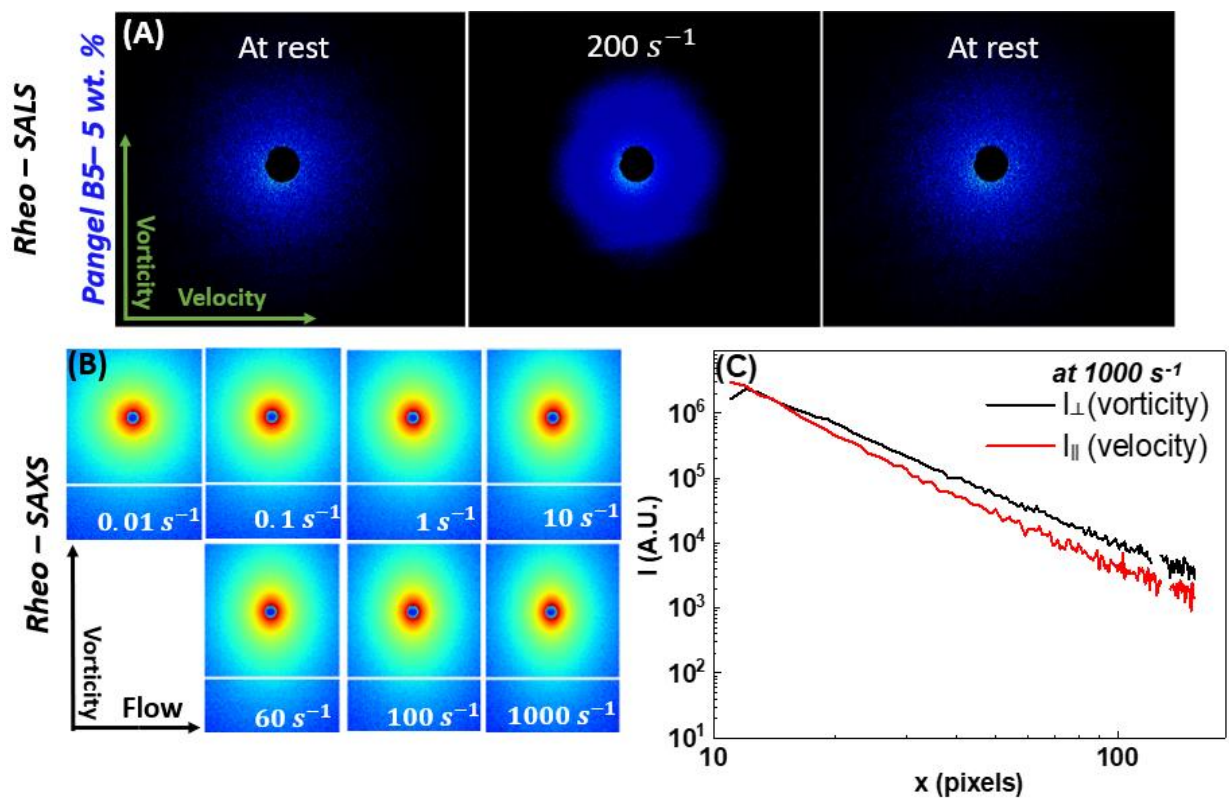

**Fig. S7:** Pangel B5 dispersion ($c = 5wt.\%$) under steady shear. Evolution of: (A) 2D-SALS at rest, at shear rate 200 s$^{-1}$ and at rest immediately after flow cessation (B) 2D-SAXS patterns obtained from high to low shear rates. (C) SAXS Intensity profiles, $I(q)$ in flow and vorticity plane direction at 1000 s$^{-1}$.

***Section IV:*** *Tuning of clay dispersions through steady and oscillatory shear*

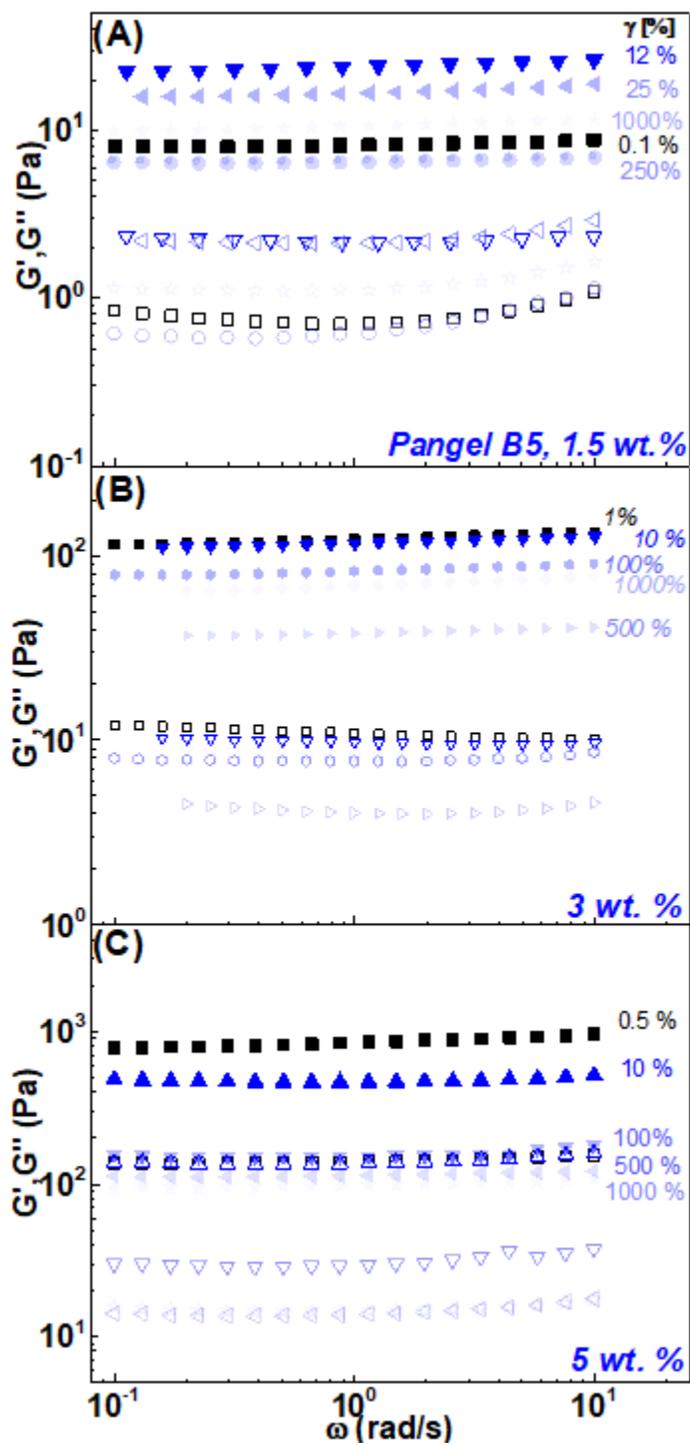

**Fig. S8:** Frequency dependence storage, $G'$ (filled) and loss modulus, $G''$ (open symbols) at $\gamma = 0.1\%, \omega = 1\ rad/s$ for different concentrations of Pangel B5 samples at (A) 1.5 wt. %, (B) 3 wt. % and

(C) 5 wt. %. Measurements were performed from high to low frequencies after preparation of the gels at different oscillatory strains as depicted in the legends.

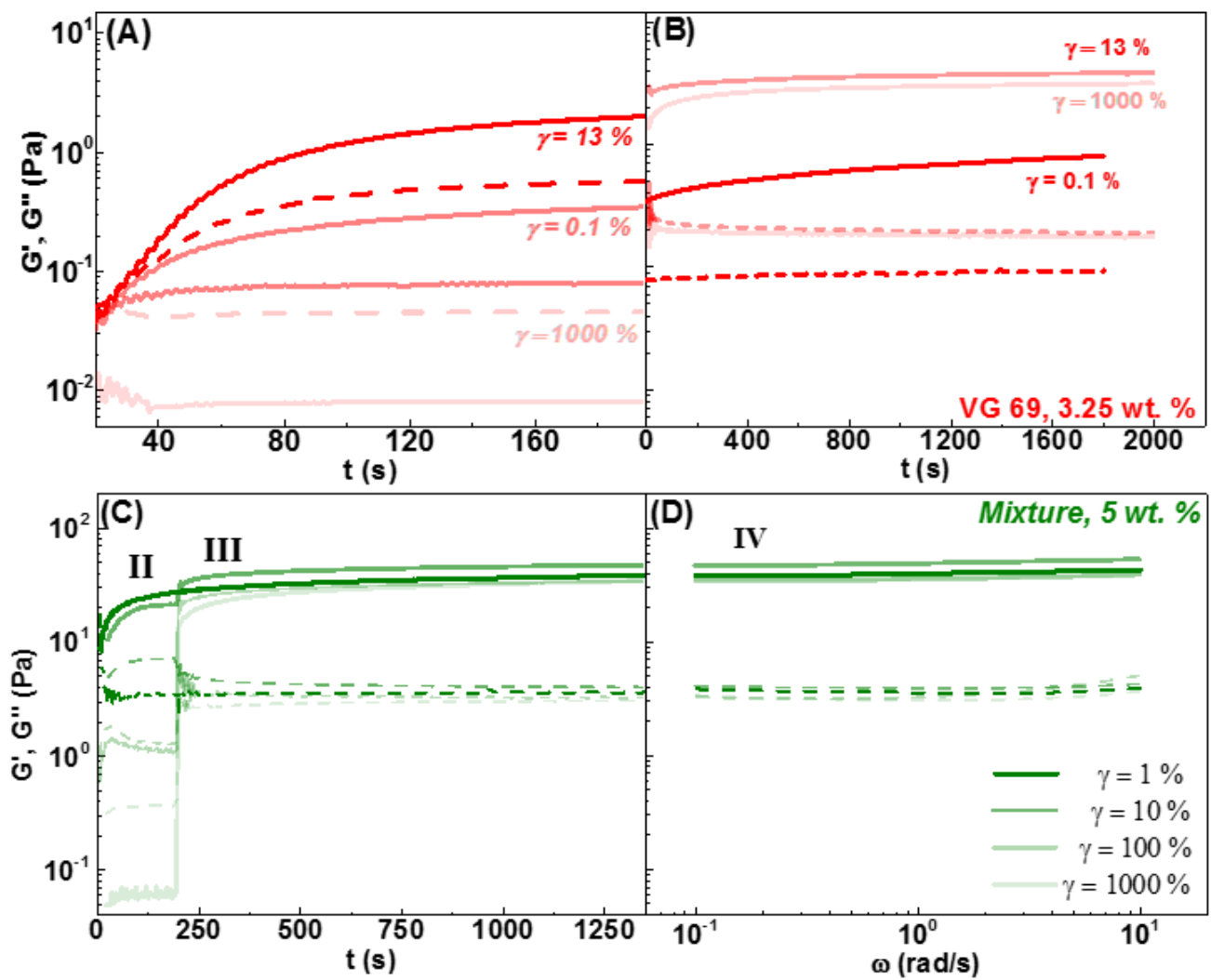

**Fig. S9:** Storage modulus $G'$ (solid lines) and loss modulus, $G''$ (dashed lines) in different regimes during the shear-tuning protocol. (A) Moduli evolving with time in period II ( $0.1 < \gamma < 1000\%$ from light red to red lines); (B) period III (upon flow cessation, from $\gamma = \gamma_{legend}$ to $\gamma = 0.1\%$ ), (DTS at $\omega = 1 \, rad/s$) for a VG-69 dispersion (3.25 $wt.\%$); (C) Period II ($1 < \gamma < 1000\%$ from light green to green lines) and III (upon flow cessation from, $\gamma = \gamma_{legend}$ to $\gamma = 1\%$), (DTS at ω=1 rad/s); (D) with frequency (DFS at $\gamma = 1\%$) in period IV for a Mixture dispersion (5 wt%).

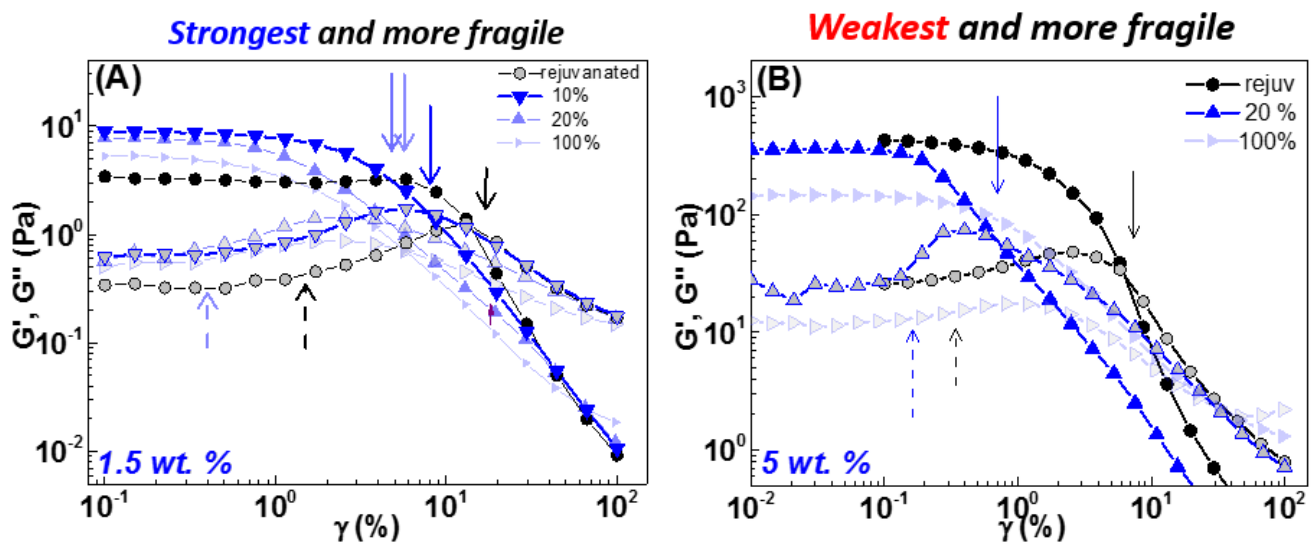

**Fig. S10:** Dynamic strain sweep measurements performed after different LAOS measurements according to the rheological protocol in the main text from: low to high strain amplitudes for Pangel B5 dispersions at (A) 1.5 and (B) 5 $wt.\%$.

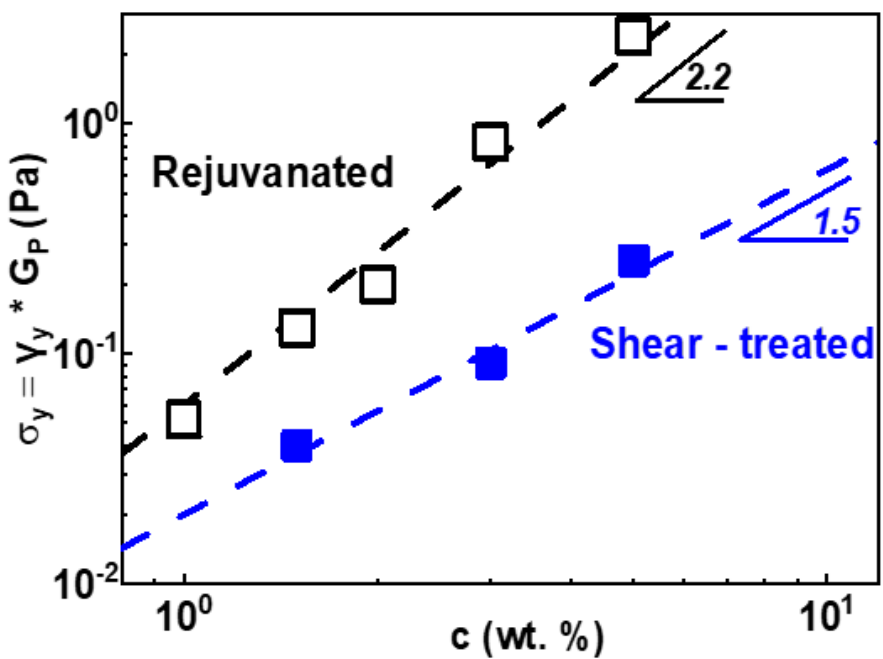

**Fig. S11:** yield stress ($\sigma_y$) deduced from yield strain (onset of nonlinearity) times the plateau modulus ($G_P$) for Pangel B5 dispersions after rejuvenation (open black squares) or prepared at different steady shear rates (filled blue).

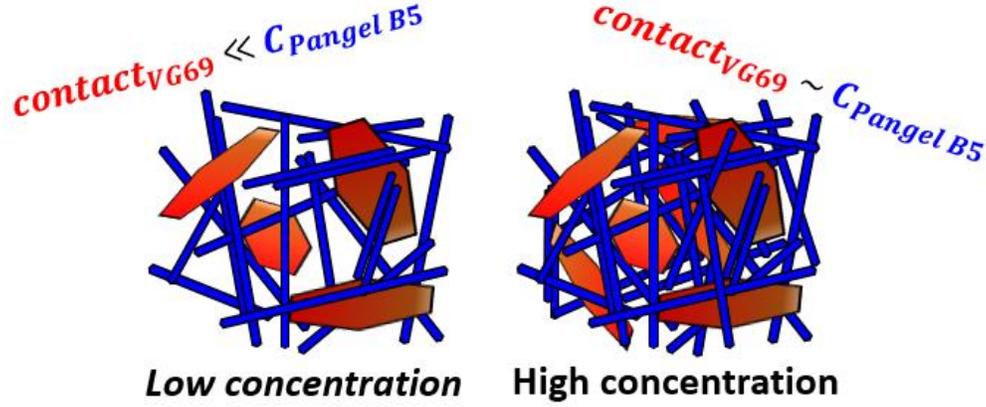

**Fig. S12:** Tentative cartoon of Pangel B5 (blue rods) and VG-69 particles at (50-50 wt. %) at low (left) and high (right) clay concentration.

***Section V:*** *Scaling behavior of the elastic properties of clay dispersions.*

|  | $G_P \sim c^A$ | $\gamma_y \sim c^B$ | *Strong –link model (a =0)* | | *Transition model* | | | SAXS |
|---|---|---|---|---|---|---|---|---|
| **System** | **A** | **B** | $d_f$ | $x$ | $d_f$ | $\alpha$ *(for $1 \leq x \leq 1.3$)* | $\beta$ | $d_f$ |
| VG-69 (Rejuv.) | 3.9 | -1 | 2.31 | -0.31 | 2.31 | $0.43 \leq a \leq 0.49$ | 2.7 | 2.5* |
| Pangel B5 (Rejuv.) | 3.9 | -1.7 | 2.09 | 0.6 | 2.09 | $0.15 \leq a \leq 0.23$ | 3.54 | 2.5 |
| Mixture (Rejuv.) | 2.8 | -0.4 | 2.16 | -0.648 | 2.16 | $0.5 \leq a \leq 0.6$ | 2.35 | - |
| Pangel B5 (shear treated) | 2.5 | -1 | 1.66 | 0 | 1.66 | $0.22 \leq a \leq 0.3$ | 3.34 | 2.4 |

**Table S1:** Power law exponents, A and B, for the concentration dependence of the elastic modulus and yield strain, and related parameters of the fractal gel elasticity model in the strong-link and transition regimes, $(d_f, x, \beta, \alpha)$ as well as fractal dimension $d_f$ from scattering experiments. * taken from [58].